%% file: autosam.tex
\documentclass[twocolumn]{autart}    

\input{2-compilation/1-packages}

\input{2-compilation/2-definitions} 

\begin{document}

\input{1-title_page/frontpage}

\input{3-main_content/1-introduction}
\input{3-main_content/2-preliminaries}
\input{3-main_content/3-adaptive_cbf}

\input{3-main_content/4-online_rls}
\input{3-main_content/5-robustness}
\input{3-main_content/7-simulation}

\input{3-main_content/8-conclusion}

\input{6-appendices/0-app_main}
\vspace*{-0.4cm}

\bibliographystyle{plain}        
\bibliography{main.bib}          


\input{3-main_content/9-biography}
\end{document}

%% file: 2-compilation/1-packages.tex
\usepackage{amsmath, amsfonts, amssymb}
\usepackage{dsfont}
\usepackage{graphicx}
\usepackage{subcaption}
\usepackage{algorithm}
\usepackage{algpseudocode}
\usepackage{siunitx}
\usepackage[dvipsnames]{xcolor}
\usepackage{bm}
\usepackage{booktabs}
\usepackage{multirow}
\usepackage{tabularx}
\usepackage{tikz}

%% file: 2-compilation/2-definitions.tex
\newcommand{\splus}{\hspace{-0.08cm}+\hspace{-0.08cm}}
\newcommand{\sminus}{\hspace{-0.08cm}-\hspace{-0.08cm}}
\newcommand{\seq}{\hspace{-0.08cm}=\hspace{-0.08cm}}
\newcommand{\sdeq}{\hspace{-0.08cm}:=\hspace{-0.08cm}}

\newtheorem{assumption}{Assumption}
\newtheorem{remark}{Remark}
\newtheorem{definition}{Definition}

\newcommand{\bR}{\mathbb{R}}
\newcommand{\bN}{\mathbb{N}}
\newcommand{\obN}{\overline{\bN}}
\newcommand{\bI}{\mathbb{I}}

\newcommand{\bone}{\mathbf{1}}
\newcommand{\bzero}{\mathbf{0}}

\newcommand{\est}{\mathtt{EST}}

\newcommand{\dist}{\mathrm{dist}}
\newcommand{\fd}{f_{\mathrm{d}}}

\newcommand{\inte}{\mathrm{int}}
\newcommand{\csl}{\mathcal{C}_{\mathrm{SL}}}
\newcommand{\csi}{\mathcal{C}_{\mathrm{SI}}}
\newcommand{\kei}{\mathcal{K}_{\mathrm{e},\infty}}
\newcommand{\ke}{\mathcal{K}_{\mathrm{e}}}

\newcommand{\cU}{\mathcal{U}}
\newcommand{\cX}{\mathcal{X}}
\newcommand{\cXs}{\mathcal{X}_{\mathrm{s}}}
\newcommand{\cXst}[1]{\mathcal{X}_{\mathrm{s},#1}}

\newcommand{\cW}{\mathcal{W}}
\newcommand{\cF}{\mathcal{F}}
\newcommand{\cS}{\mathcal{S}}
\newcommand{\cSb}[1]{\mathcal{S}_{B,#1}}
\newcommand{\bcSr}[1]{\overline{\mathcal{S}}_{\mathrm{ra},#1}}
\newcommand{\cSr}[1]{\mathcal{S}_{\mathrm{ra},#1}}

\DeclareMathOperator*{\argmin}{arg\,min}

\newcommand{\mG}{\lambda_{\min}(\Gamma)}
\newcommand{\Fr}{F_{\mathrm{r}}}
\newcommand{\vf}{v_{\mathrm{f}}}

\newcommand{\adp}{\mathtt{ADP}}
\newcommand{\cT}{\mathcal{T}}

\newcommand{\setpara}{\Theta}
\newcommand{\para}{\theta}
\newcommand{\tpara}{\para^\ast}
\newcommand{\hpara}{\hat{\para}}
\newcommand{\hpa}[1]{\hat{\para}_{#1}}

\newcommand{\ipa}[1]{\delta_{\para,#1}}
\newcommand{\epa}{e_{\para}}
\newcommand{\epat}[1]{e_{\para,#1}}

\newcommand{\be}{\varepsilon_{\para}}
\newcommand{\bet}[1]{\varepsilon_{\para,#1}}
\newcommand{\eet}[1]{\eta_{\para,#1}}

\newcommand{\Brt}[1]{B_{\mathrm{ra},#1}}
\newcommand{\BBrt}[1]{\overline{B}_{\mathrm{ra},#1}}

\newcommand{\iasf}[1]{u_{\mathrm{ras},#1}}

\newcommand{\Usf}{\mathcal{U}_{\mathrm{rs},\para^\ast}}
\newcommand{\uasf}[1]{\pi_{\mathrm{ras},#1}}
\newcommand{\Uasf}[1]{\mathcal{U}_{\mathrm{ras},#1}}
\newcommand{\BUasf}[1]{\overline{\mathcal{U}}_{\mathrm{ras},#1}}

\newcommand{\id}{i_{\mathrm{d}}}
\newcommand{\iq}{i_{\mathrm{q}}}
\newcommand{\uq}{u_{\mathrm{q}}}
\newcommand{\iqt}[1]{i_{\mathrm{q},#1}}
\newcommand{\np}{n_{\mathrm{p}}}
\newcommand{\phif}{\phi_{\text{f}}}

\newcommand{\uqt}[1]{u_{\mathrm{q},#1}}
\newcommand{\Bvis}{b_{\text{vis}}}

\newcommand{\mnt}[1]{{\color{CadetBlue}#1}}

%% file: 1-title_page/frontpage.tex
\begin{frontmatter}

\title{Robust Adaptive Discrete-Time Control Barrier Certificate \thanksref{footnoteinfo}} 

\thanks[footnoteinfo]{This paper is part of a project that has received funding from the 
European Research Council (ERC) under the European Union’s Horizon 
2020 research and innovation programme (Grant agreement No. 101018826 - CLariNet), and it was not presented at any IFAC meeting. Corresponding authors: Changrui Liu \& Shengling Shi.}

\author[tud]{Changrui Liu}\ead{c.liu-14@tudelft.nl},    
\author[tud]{Anil Alan}\ead{a.alan@tudelft.nl},               
\author[tud]{Shengling Shi}\ead{s.shi-3@tudelft.nl},
\author[tud]{Bart De Schutter}\ead{b.deschutter@tudelft.nl}  

\address[tud]{Delft Center for Systems and Control, TU Delft, Delft, The Netherlands.}  


\input{3-main_content/0-abstract}

\end{frontmatter}

%% file: 3-main_content/0-abstract.tex
\begin{abstract}                          
This work develops a robust adaptive control strategy for discrete-time systems using Control Barrier Functions (CBFs) to ensure safety under parametric model uncertainty and disturbances. A key contribution of this work is establishing a barrier function certificate in discrete time for general online parameter estimation algorithms. This barrier function certificate guarantees positive invariance of the safe set despite disturbances and parametric uncertainty without access to the true system parameters. In addition, real-time implementation and inherent robustness guarantees are provided. The proposed robust adaptive safe control framework demonstrates that the parameter estimation module can be designed separately from the CBF-based safety filter, simplifying the development of safe adaptive controllers for discrete-time systems. The resulting safe control approach guarantees that the system remains within the safe set while the controller adapts to model uncertainties, making it a promising strategy for discrete-time safety-critical systems.
\end{abstract}

%% file: 3-main_content/1-introduction.tex
\section{Introduction}
\vspace*{-0.4cm}
Safety-critical control aims to guarantee that the system state remains within a safe set, i.e., constraint satisfaction, which is crucial in domains where unsafe behaviors can lead to undesirable outcomes \cite{ames2016control}. Among various approaches (e.g., Hamilton-Jacobi reachability analysis~\cite{semnani2020multi} and Model Predictive Control (MPC) \cite{liu2026certainty}, \cite{rawlings2017model}), Control Barrier Functions (CBFs) have become a powerful and systematic tool for both continuous-time \cite{ames2016control} and discrete-time systems \cite{agrawal2017discrete}. CBFs allow formal encoding of safety as control invariant sets, and they can be integrated on top of traditional control methods (e.g., control Lyapunov functions~\cite{ames2016control}) and learning-based control approaches (e.g., model-based Reinforcement Learning (RL)~\cite{cohen2023safe}) that may not inherently offer safety guarantees. CBFs have many applications, including control of bipedal robots~\cite{agrawal2017discrete}, cruise control~\cite{ames2016control}, and microgrid energy management~\cite{liu2025approximate}, demonstrating their versatility and practical relevance in safety-critical domains.

Despite the recent surge in learning-based approaches for CBFs~\cite{srinivasan2020synthesis}, model-based methods remain essential due to their analytical tractability and formal safety guarantees~\cite{ames2016control}, \cite{wang2025safe}. Typically, these approaches support modular design through the explicit use of dynamical models \cite{agrawal2017discrete}, \cite{ames2016control}, \cite{xu2015robustness}. A standard approach is to encode safety constraints through a CBF-based inequality condition on the input, and then to incorporate this condition within a safety filter~\cite{didier2024approximate} that minimizes the adjustment of a given nominal input that may otherwise lead to unsafe behaviors (i.e., constraint violations). However, a fundamental challenge in model-based methods is model uncertainty, which can compromise safety \cite{wang2025safe}. Accordingly, various robust and adaptive extensions have been developed for CBFs to 
handle disturbances~\cite{cosner2024bounding}, \cite{jankovic2018robust} or parametric uncertainty~\cite{lopez2020robust}, \cite{shen2025adaptive}, \cite{taylor2020adaptive}. For disturbances, high-probability safety~\cite{cosner2024bounding} and guaranteed safety~\cite{jankovic2018robust} have been established for unbounded and bounded disturbances, respectively. There are also results on adaptive safety focusing on parametric uncertainty, employing either Lyapunov-based methods~\cite{lopez2020robust}, \cite{taylor2020adaptive}, or batched least-squares methods \cite{shen2025adaptive}. For an in-depth overview, we refer the reader to \cite{wang2025safe}.

\vspace*{-0.2cm}
Although recent progress has been made on robust and adaptive CBFs in continuous time~\cite{lopez2020robust}, \cite{shen2025adaptive}, \cite{taylor2020adaptive}, the discrete-time case remains largely unexplored despite its relevance to discrete-time control designs (e.g., approximate MPC~\cite{hertneck2018learning}, \cite{liu2025approximate} and model-based RL~\cite{cohen2023safe}). Since practical systems rely on sampled sensing and actuation, there is a need to develop adaptive Discrete-Time CBFs (DT-CBFs)~\cite{cosner2024bounding}, \cite{santoyo2019verification}. The main difficulty for DT-CBFs is that derivative-based arguments are unavailable. In addition, \textit{causal} controllers must compute inputs without future information, a feature overlooked in continuous-time methods~\cite{lopez2020robust}, \cite{shen2025adaptive}. Moreover, Lyapunov adaptation laws lack guarantees on bounded updates or convergence~\cite{lopez2020robust}, motivating estimation-based approaches that better align with sampled-data settings~\cite{cohen2023safe}. Since standard estimation tools (e.g., least-squares methods) operate on discrete measurements~\cite{ljung2010perspectives}, DT-CBFs provide a more suitable foundation to establish adaptive safety.

\vspace*{-0.3cm}
This paper extends adaptive safe control to discrete-time systems, particularly those implemented
in embedded or sampled-data settings. The proposed method connects to classical adaptive control, where parameter estimation, projection, and set-membership identification are used to ensure robustness~\cite{astrom1994adaptive}, \cite{goodwin2014adaptive}, \cite{ioannou1996robust}, \cite{milanese2013bounding}. In contrast, this paper reverses the standard perspective: rather than augmenting adaptive laws with safety mechanisms, robust safety is prioritized as the primary objective, and adaptation is subsequently integrated via parameter estimation to reduce conservatism. The contributions of this work are summarized as follows:
\vspace*{-0.3cm}
\begin{itemize}
	\item Given a DT-CBF satisfying appropriate conditions, we establish safety guarantees for input-affine discrete-time systems under bounded disturbances and parametric mismatch, thereby embedding robust adaptive DT-CBFs into a practically relevant discrete-time control framework.
	\item A modular control architecture is proposed in which parameter estimation is decoupled from CBF-based safety filtering, in contrast to existing approaches~\cite{lopez2020robust}, \cite{taylor2020adaptive} where the adaptation law is co-designed with the safety filter. The proposed framework is compatible with any nominal controller and any parameter estimator with a bounded update rate, thereby significantly broadening applicability.
	\item The \textit{inherent robustness}\hspace*{-0.2em}\footnote{In this paper, inherent robustness refers to steering the state towards safe sets when the current state is unsafe \cite[Section 2.2]{xu2015robustness}, which should not be confused with robustness against model uncertainty \cite{didier2024approximate}, \cite{lopez2020robust}.}of the proposed safe control approach is rigorously analyzed, extending results for continuous-time systems without model uncertainty \cite{xu2015robustness} to discrete-time uncertain systems.
\end{itemize}

\vspace*{-0.3cm}
It is further noted that the primary contribution of this paper lies in establishing a general robust adaptive safe control framework with a given DT-CBF, rather than designing DT-CBFs, developing novel estimation algorithms, or refining the safety filter framework~\cite{didier2024approximate}. The rest of the paper is organized as follows. Section~\ref{sec2:preliminaries} provides preliminaries, and Section~\ref{sec3:robust_adaptive_safety} introduces robust adaptive DT-CBFs. A simple estimation algorithm is given in Section~\ref{sec4:identification}. Inherent robustness guarantees are discussed in Section~\ref{sec5:robustness}. A simulation example is presented in Section~\ref{sec7:numerical_example}, and~Section \ref{sec8:conclusion} concludes the paper and outlines future research directions.
\input{5-figures/fig_diagram}

%% file: 5-figures/fig_diagram.tex
\begin{figure}[h]
	\centering 
	\resizebox{\columnwidth}{!}{%

	\tikzset{every picture/.style={line width=0.75pt}} 
	
	\begin{tikzpicture}[x=0.75pt,y=0.75pt,yscale=-1,xscale=1]
		
		\draw  [color={rgb, 255:red, 80; green, 227; blue, 194 }  ,draw opacity=1 ][fill={rgb, 255:red, 80; green, 227; blue, 194 }  ,fill opacity=0.2 ][dash pattern={on 4.5pt off 4.5pt}] (28.2,50.4) .. controls (28.2,47.64) and (30.44,45.4) .. (33.2,45.4) -- (253.2,45.4) .. controls (255.96,45.4) and (258.2,47.64) .. (258.2,50.4) -- (258.2,232.6) .. controls (258.2,235.36) and (255.96,237.6) .. (253.2,237.6) -- (33.2,237.6) .. controls (30.44,237.6) and (28.2,235.36) .. (28.2,232.6) -- cycle ;
		\draw  [color={rgb, 255:red, 184; green, 233; blue, 134 }  ,draw opacity=1 ][fill={rgb, 255:red, 126; green, 211; blue, 33 }  ,fill opacity=0.2 ][line width=0.75]  (329.2,85.2) .. controls (329.2,80.45) and (333.05,76.6) .. (337.8,76.6) -- (489.6,76.6) .. controls (494.35,76.6) and (498.2,80.45) .. (498.2,85.2) -- (498.2,111) .. controls (498.2,115.75) and (494.35,119.6) .. (489.6,119.6) -- (337.8,119.6) .. controls (333.05,119.6) and (329.2,115.75) .. (329.2,111) -- cycle ;
		\draw  [fill={rgb, 255:red, 155; green, 155; blue, 155 }  ,fill opacity=0.2 ] (517,93.3) .. controls (517,89.82) and (519.73,87) .. (523.1,87) .. controls (526.47,87) and (529.2,89.82) .. (529.2,93.3) .. controls (529.2,96.78) and (526.47,99.6) .. (523.1,99.6) .. controls (519.73,99.6) and (517,96.78) .. (517,93.3) -- cycle ; \draw   (518.79,88.85) -- (527.41,97.75) ; \draw   (527.41,88.85) -- (518.79,97.75) ;
		\draw [color={rgb, 255:red, 65; green, 117; blue, 5 }  ,draw opacity=1 ]   (498.2,92.6) -- (514,92.94) ;
		\draw [shift={(517,93)}, rotate = 181.22] [fill={rgb, 255:red, 65; green, 117; blue, 5 }  ,fill opacity=1 ][line width=0.08]  [draw opacity=0] (8.93,-4.29) -- (0,0) -- (8.93,4.29) -- cycle    ;
		\draw    (523.2,60.6) -- (523.47,84) ;
		\draw [shift={(523.5,87)}, rotate = 269.35] [fill={rgb, 255:red, 0; green, 0; blue, 0 }  ][line width=0.08]  [draw opacity=0] (8.93,-4.29) -- (0,0) -- (8.93,4.29) -- cycle    ;
		\draw [color={rgb, 255:red, 65; green, 117; blue, 5 }  ,draw opacity=1 ]   (288.2,93.6) -- (325.2,93.6) ;
		\draw [shift={(328.2,93.6)}, rotate = 180] [fill={rgb, 255:red, 65; green, 117; blue, 5 }  ,fill opacity=1 ][line width=0.08]  [draw opacity=0] (8.93,-4.29) -- (0,0) -- (8.93,4.29) -- cycle    ;
		\draw [color={rgb, 255:red, 245; green, 166; blue, 35 }  ,draw opacity=1 ][line width=1.5]    (313.2,105.6) -- (324.2,105.6) ;
		\draw [shift={(328.2,105.6)}, rotate = 180] [fill={rgb, 255:red, 245; green, 166; blue, 35 }  ,fill opacity=1 ][line width=0.08]  [draw opacity=0] (9.29,-4.46) -- (0,0) -- (9.29,4.46) -- cycle    ;
		\draw [color={rgb, 255:red, 65; green, 117; blue, 5 }  ,draw opacity=1 ]   (529,93) -- (558.2,93.54) ;
		\draw [shift={(561.2,93.6)}, rotate = 181.07] [fill={rgb, 255:red, 65; green, 117; blue, 5 }  ,fill opacity=1 ][line width=0.08]  [draw opacity=0] (8.93,-4.29) -- (0,0) -- (8.93,4.29) -- cycle    ;
		\draw [color={rgb, 255:red, 245; green, 166; blue, 35 }  ,draw opacity=1 ][line width=1.5]    (266,106) -- (266,148.6) ;
		\draw [color={rgb, 255:red, 65; green, 117; blue, 5 }  ,draw opacity=1 ]   (308,93) -- (308.2,158.6) ;
		\draw [color={rgb, 255:red, 245; green, 166; blue, 35 }  ,draw opacity=1 ][line width=1.5]    (314.2,147.6) -- (347.2,147.6) ;
		\draw [shift={(351.2,147.6)}, rotate = 180] [fill={rgb, 255:red, 245; green, 166; blue, 35 }  ,fill opacity=1 ][line width=0.08]  [draw opacity=0] (9.29,-4.46) -- (0,0) -- (9.29,4.46) -- cycle    ;
		\draw [color={rgb, 255:red, 65; green, 117; blue, 5 }  ,draw opacity=1 ]   (308.2,159.6) -- (348.2,159.6) ;
		\draw [shift={(351.2,159.6)}, rotate = 180] [fill={rgb, 255:red, 65; green, 117; blue, 5 }  ,fill opacity=1 ][line width=0.08]  [draw opacity=0] (8.93,-4.29) -- (0,0) -- (8.93,4.29) -- cycle    ;
		\draw  [color={rgb, 255:red, 74; green, 144; blue, 226 }  ,draw opacity=1 ][fill={rgb, 255:red, 74; green, 144; blue, 226 }  ,fill opacity=0.2 ][line width=0.75]  (352,142) .. controls (352,137.58) and (355.58,134) .. (360,134) -- (478.2,134) .. controls (482.62,134) and (486.2,137.58) .. (486.2,142) -- (486.2,166) .. controls (486.2,170.42) and (482.62,174) .. (478.2,174) -- (360,174) .. controls (355.58,174) and (352,170.42) .. (352,166) -- cycle ;
		\draw [color={rgb, 255:red, 65; green, 117; blue, 5 }  ,draw opacity=1 ]   (536,93) -- (536.2,152.6) ;
		\draw [color={rgb, 255:red, 65; green, 117; blue, 5 }  ,draw opacity=1 ][fill={rgb, 255:red, 0; green, 0; blue, 0 }  ,fill opacity=1 ]   (536.2,152.6) -- (489.2,152.6) ;
		\draw [shift={(486.2,152.6)}, rotate = 360] [fill={rgb, 255:red, 65; green, 117; blue, 5 }  ,fill opacity=1 ][line width=0.08]  [draw opacity=0] (8.93,-4.29) -- (0,0) -- (8.93,4.29) -- cycle    ;
		\draw [color={rgb, 255:red, 74; green, 144; blue, 226 }  ,draw opacity=1 ][line width=1.5]    (431,174) -- (431.2,202.6) ;
		\draw [color={rgb, 255:red, 74; green, 144; blue, 226 }  ,draw opacity=1 ][line width=1.5]    (241.4,200.21) -- (410.2,200.6) ;
		\draw [shift={(237.4,200.2)}, rotate = 0.13] [fill={rgb, 255:red, 74; green, 144; blue, 226 }  ,fill opacity=1 ][line width=0.08]  [draw opacity=0] (9.29,-4.46) -- (0,0) -- (9.29,4.46) -- cycle    ;
		\draw  [color={rgb, 255:red, 139; green, 87; blue, 42 }  ,draw opacity=1 ][fill={rgb, 255:red, 139; green, 87; blue, 42 }  ,fill opacity=0.2 ][line width=0.75]  (97.2,194) .. controls (97.2,189.58) and (100.78,186) .. (105.2,186) -- (228.2,186) .. controls (232.62,186) and (236.2,189.58) .. (236.2,194) -- (236.2,218) .. controls (236.2,222.42) and (232.62,226) .. (228.2,226) -- (105.2,226) .. controls (100.78,226) and (97.2,222.42) .. (97.2,218) -- cycle ;
		\draw [color={rgb, 255:red, 74; green, 144; blue, 226 }  ,draw opacity=1 ][line width=1.5]    (415.03,174.1) -- (414.83,215.5) ;
		\draw [color={rgb, 255:red, 74; green, 144; blue, 226 }  ,draw opacity=1 ][line width=1.5]    (415,215) -- (241.2,214.61) ;
		\draw [shift={(237.2,214.6)}, rotate = 0.13] [fill={rgb, 255:red, 74; green, 144; blue, 226 }  ,fill opacity=1 ][line width=0.08]  [draw opacity=0] (9.29,-4.46) -- (0,0) -- (9.29,4.46) -- cycle    ;
		\draw  [color={rgb, 255:red, 208; green, 2; blue, 27 }  ,draw opacity=0.4 ][fill={rgb, 255:red, 208; green, 2; blue, 27 }  ,fill opacity=0.2 ][line width=0.75]  (88.2,143) .. controls (88.2,138.58) and (91.78,135) .. (96.2,135) -- (237,135) .. controls (241.42,135) and (245,138.58) .. (245,143) -- (245,167) .. controls (245,171.42) and (241.42,175) .. (237,175) -- (96.2,175) .. controls (91.78,175) and (88.2,171.42) .. (88.2,167) -- cycle ;
		\draw  [color={rgb, 255:red, 245; green, 166; blue, 35 }  ,draw opacity=1 ][fill={rgb, 255:red, 245; green, 166; blue, 35 }  ,fill opacity=0.2 ][line width=0.75]  (115,94) .. controls (115,89.58) and (118.58,86) .. (123,86) -- (217,86) .. controls (221.42,86) and (225,89.58) .. (225,94) -- (225,118) .. controls (225,122.42) and (221.42,126) .. (217,126) -- (123,126) .. controls (118.58,126) and (115,122.42) .. (115,118) -- cycle ;
		\draw [color={rgb, 255:red, 208; green, 2; blue, 27 }  ,draw opacity=1 ]   (61,102) -- (61.2,156.6) ;
		\draw [color={rgb, 255:red, 139; green, 87; blue, 42 }  ,draw opacity=1 ]   (73.63,113.17) -- (73.83,199.77) ;
		\draw [color={rgb, 255:red, 65; green, 117; blue, 5 }  ,draw opacity=1 ]   (287.9,56.2) -- (288.2,93.6) ;
		\draw [color={rgb, 255:red, 65; green, 117; blue, 5 }  ,draw opacity=1 ]   (287.9,56.2) -- (394.5,56) ;
		\draw  [fill={rgb, 255:red, 155; green, 155; blue, 155 }  ,fill opacity=0.2 ] (395,45.6) .. controls (395,43.94) and (396.34,42.6) .. (398,42.6) -- (417.9,42.6) .. controls (419.56,42.6) and (420.9,43.94) .. (420.9,45.6) -- (420.9,65.5) .. controls (420.9,67.16) and (419.56,68.5) .. (417.9,68.5) -- (398,68.5) .. controls (396.34,68.5) and (395,67.16) .. (395,65.5) -- cycle ;
		\draw [color={rgb, 255:red, 65; green, 117; blue, 5 }  ,draw opacity=1 ]   (423.9,56.19) -- (539.5,56) ;
		\draw [shift={(420.9,56.2)}, rotate = 359.9] [fill={rgb, 255:red, 65; green, 117; blue, 5 }  ,fill opacity=1 ][line width=0.08]  [draw opacity=0] (8.93,-4.29) -- (0,0) -- (8.93,4.29) -- cycle    ;
		\draw [color={rgb, 255:red, 65; green, 117; blue, 5 }  ,draw opacity=1 ]   (539.5,56) -- (539.8,93.4) ;
		\draw [color={rgb, 255:red, 139; green, 87; blue, 42 }  ,draw opacity=1 ]   (73.13,113.17) -- (111.13,113.17) ;
		\draw [shift={(114.13,113.17)}, rotate = 180] [fill={rgb, 255:red, 139; green, 87; blue, 42 }  ,fill opacity=1 ][line width=0.08]  [draw opacity=0] (8.93,-4.29) -- (0,0) -- (8.93,4.29) -- cycle    ;
		\draw [color={rgb, 255:red, 74; green, 144; blue, 226 }  ,draw opacity=1 ][line width=1.5]  [dash pattern={on 5.63pt off 4.5pt}]  (285.2,155.6) -- (285,200) ;
		\draw [color={rgb, 255:red, 208; green, 2; blue, 27 }  ,draw opacity=1 ][fill={rgb, 255:red, 0; green, 0; blue, 0 }  ,fill opacity=1 ]   (111,102) -- (61,102) ;
		\draw [shift={(114,102)}, rotate = 180] [fill={rgb, 255:red, 208; green, 2; blue, 27 }  ,fill opacity=1 ][line width=0.08]  [draw opacity=0] (8.93,-4.29) -- (0,0) -- (8.93,4.29) -- cycle    ;
		\draw [color={rgb, 255:red, 139; green, 87; blue, 42 }  ,draw opacity=1 ]   (73.63,200.17) -- (97.63,200.17) ;
		\draw [color={rgb, 255:red, 208; green, 2; blue, 27 }  ,draw opacity=1 ][fill={rgb, 255:red, 0; green, 0; blue, 0 }  ,fill opacity=1 ]   (88.2,156.6) -- (78.2,156.6) ;
		\draw  [draw opacity=0] (67.99,156.37) .. controls (68.07,153.51) and (70.34,151.21) .. (73.11,151.22) .. controls (75.93,151.23) and (78.21,153.61) .. (78.2,156.53) .. controls (78.2,156.56) and (78.2,156.58) .. (78.2,156.6) -- (73.1,156.52) -- cycle ; \draw  [color={rgb, 255:red, 208; green, 2; blue, 27 }  ,draw opacity=1 ] (67.99,156.37) .. controls (68.07,153.51) and (70.34,151.21) .. (73.11,151.22) .. controls (75.93,151.23) and (78.21,153.61) .. (78.2,156.53) .. controls (78.2,156.56) and (78.2,156.58) .. (78.2,156.6) ;  
		\draw [color={rgb, 255:red, 208; green, 2; blue, 27 }  ,draw opacity=1 ][fill={rgb, 255:red, 0; green, 0; blue, 0 }  ,fill opacity=1 ]   (68.2,156.6) -- (61.2,156.6) ;
		\draw [color={rgb, 255:red, 74; green, 144; blue, 226 }  ,draw opacity=1 ][line width=1.5]  [dash pattern={on 5.63pt off 4.5pt}]  (249.2,153.6) -- (285.2,153.6) ;
		\draw [shift={(245.2,153.6)}, rotate = 0] [fill={rgb, 255:red, 74; green, 144; blue, 226 }  ,fill opacity=1 ][line width=0.08]  [draw opacity=0] (9.29,-4.46) -- (0,0) -- (9.29,4.46) -- cycle    ;
		\draw  [draw opacity=0][line width=1.5]  (409.7,201.57) .. controls (409.7,201.57) and (409.7,201.57) .. (409.7,201.57) .. controls (409.7,201.57) and (409.7,201.57) .. (409.7,201.57) .. controls (409.71,198.65) and (412.01,196.29) .. (414.83,196.31) .. controls (417.65,196.32) and (419.92,198.7) .. (419.91,201.63) -- (414.8,201.6) -- cycle ; \draw  [color={rgb, 255:red, 74; green, 144; blue, 226 }  ,draw opacity=1 ][line width=1.5]  (409.7,201.57) .. controls (409.7,201.57) and (409.7,201.57) .. (409.7,201.57) .. controls (409.7,201.57) and (409.7,201.57) .. (409.7,201.57) .. controls (409.71,198.65) and (412.01,196.29) .. (414.83,196.31) .. controls (417.65,196.32) and (419.92,198.7) .. (419.91,201.63) ;  
		\draw [color={rgb, 255:red, 74; green, 144; blue, 226 }  ,draw opacity=1 ][line width=1.5]    (419.4,201.2) -- (431.2,201.6) ;
		\draw  [draw opacity=0][line width=1.5]  (302.99,148.55) .. controls (302.99,148.55) and (302.99,148.55) .. (302.99,148.55) .. controls (302.99,148.55) and (302.99,148.55) .. (302.99,148.55) .. controls (303.01,145.62) and (305.3,143.26) .. (308.12,143.28) .. controls (310.94,143.29) and (313.22,145.68) .. (313.2,148.6) -- (308.1,148.57) -- cycle ; \draw  [color={rgb, 255:red, 245; green, 166; blue, 35 }  ,draw opacity=1 ][line width=1.5]  (302.99,148.55) .. controls (302.99,148.55) and (302.99,148.55) .. (302.99,148.55) .. controls (302.99,148.55) and (302.99,148.55) .. (302.99,148.55) .. controls (303.01,145.62) and (305.3,143.26) .. (308.12,143.28) .. controls (310.94,143.29) and (313.22,145.68) .. (313.2,148.6) ;  
		\draw [color={rgb, 255:red, 245; green, 166; blue, 35 }  ,draw opacity=1 ][line width=1.5]    (266,147.6) -- (303.2,147.6) ;
		\draw  [draw opacity=0][line width=1.5]  (302.99,106.55) .. controls (302.99,106.55) and (302.99,106.55) .. (302.99,106.55) .. controls (302.99,106.55) and (302.99,106.55) .. (302.99,106.55) .. controls (303.01,103.62) and (305.3,101.26) .. (308.12,101.28) .. controls (310.94,101.29) and (313.22,103.68) .. (313.2,106.6) -- (308.1,106.57) -- cycle ; \draw  [color={rgb, 255:red, 245; green, 166; blue, 35 }  ,draw opacity=1 ][line width=1.5]  (302.99,106.55) .. controls (302.99,106.55) and (302.99,106.55) .. (302.99,106.55) .. controls (302.99,106.55) and (302.99,106.55) .. (302.99,106.55) .. controls (303.01,103.62) and (305.3,101.26) .. (308.12,101.28) .. controls (310.94,101.29) and (313.22,103.68) .. (313.2,106.6) ;  
		\draw [color={rgb, 255:red, 245; green, 166; blue, 35 }  ,draw opacity=1 ][line width=1.5]    (225.2,105.6) -- (304.2,105.6) ;
		\draw  [color={rgb, 255:red, 189; green, 16; blue, 224 }  ,draw opacity=0.2 ][fill={rgb, 255:red, 144; green, 19; blue, 254 }  ,fill opacity=0.11 ][dash pattern={on 4.5pt off 4.5pt}] (100,259.6) .. controls (100,254.63) and (104.03,250.6) .. (109,250.6) -- (435.2,250.6) .. controls (440.17,250.6) and (444.2,254.63) .. (444.2,259.6) -- (444.2,286.6) .. controls (444.2,291.57) and (440.17,295.6) .. (435.2,295.6) -- (109,295.6) .. controls (104.03,295.6) and (100,291.57) .. (100,286.6) -- cycle ;
		\draw [color={rgb, 255:red, 74; green, 144; blue, 226 }  ,draw opacity=1 ]   (465.5,174) -- (465.8,265.4) ;
		\draw [color={rgb, 255:red, 74; green, 144; blue, 226 }  ,draw opacity=1 ]   (447.8,265.4) -- (465.8,265.4) ;
		\draw [shift={(444.8,265.4)}, rotate = 0] [fill={rgb, 255:red, 74; green, 144; blue, 226 }  ,fill opacity=1 ][line width=0.08]  [draw opacity=0] (8.93,-4.29) -- (0,0) -- (8.93,4.29) -- cycle    ;
		\draw [color={rgb, 255:red, 0; green, 0; blue, 0 }  ,draw opacity=1 ]   (63.8,280.4) -- (97.2,280.58) ;
		\draw [shift={(100.2,280.6)}, rotate = 180.31] [fill={rgb, 255:red, 0; green, 0; blue, 0 }  ,fill opacity=1 ][line width=0.08]  [draw opacity=0] (8.93,-4.29) -- (0,0) -- (8.93,4.29) -- cycle    ;
		\draw [color={rgb, 255:red, 65; green, 117; blue, 5 }  ,draw opacity=1 ]   (514.2,105.6) -- (514.4,283.2) ;
		\draw [color={rgb, 255:red, 65; green, 117; blue, 5 }  ,draw opacity=1 ]   (447.2,282.6) -- (514.2,282.6) ;
		\draw [shift={(444.2,282.6)}, rotate = 0] [fill={rgb, 255:red, 65; green, 117; blue, 5 }  ,fill opacity=1 ][line width=0.08]  [draw opacity=0] (8.93,-4.29) -- (0,0) -- (8.93,4.29) -- cycle    ;
		\draw [color={rgb, 255:red, 65; green, 117; blue, 5 }  ,draw opacity=1 ]   (498.2,105.6) -- (514.2,105.6) ;
		\draw [color={rgb, 255:red, 189; green, 16; blue, 224 }  ,draw opacity=1 ][line width=1.5]    (167.03,230) -- (167.2,250.6) ;
		\draw [shift={(167,226)}, rotate = 89.53] [fill={rgb, 255:red, 189; green, 16; blue, 224 }  ,fill opacity=1 ][line width=0.08]  [draw opacity=0] (9.29,-4.46) -- (0,0) -- (9.29,4.46) -- cycle    ;
		\draw [color={rgb, 255:red, 0; green, 0; blue, 0 }  ,draw opacity=1 ]   (63.8,265.4) -- (97.2,265.58) ;
		\draw [shift={(100.2,265.6)}, rotate = 180.31] [fill={rgb, 255:red, 0; green, 0; blue, 0 }  ,fill opacity=1 ][line width=0.08]  [draw opacity=0] (8.93,-4.29) -- (0,0) -- (8.93,4.29) -- cycle    ;
		\draw    (74.2,213.6) -- (74.5,266) ;
		\draw [color={rgb, 255:red, 0; green, 0; blue, 0 }  ,draw opacity=1 ]   (74.2,213.6) -- (93.6,213.77) ;
		\draw [shift={(96.6,213.8)}, rotate = 180.51] [fill={rgb, 255:red, 0; green, 0; blue, 0 }  ,fill opacity=1 ][line width=0.08]  [draw opacity=0] (8.93,-4.29) -- (0,0) -- (8.93,4.29) -- cycle    ;
		
		\draw (499,60.4) node [anchor=north west][inner sep=0.75pt]  [font=\large]  {$w_{t}$};
		\draw (267,68.4) node [anchor=north west][inner sep=0.75pt]  [font=\large,color={rgb, 255:red, 65; green, 117; blue, 5 }  ,opacity=1 ]  {$x{_{t}}$};
		\draw (548,71.4) node [anchor=north west][inner sep=0.75pt]  [font=\large,color={rgb, 255:red, 65; green, 117; blue, 5 }  ,opacity=1 ]  {$x_{t+1}$};
		\draw (268,108) node [anchor=north west][inner sep=0.75pt]  [font=\large,color={rgb, 255:red, 245; green, 166; blue, 35 }  ,opacity=1 ]  {$u_{\text{ras} ,t}$};
		\draw (362,150) node [anchor=north west][inner sep=0.75pt]  [font=\normalsize]  {$\est$};
		\draw (391,147) node [anchor=north west][inner sep=0.75pt]  [font=\normalsize] [align=left] {\eqref{eq:sec4-non_falsified_set}--\eqref{eq:sec4-sm_projection}, $\bar{\delta}_{\theta}$};
		\draw (319,171.4) node [anchor=north west][inner sep=0.75pt]  [font=\large,color={rgb, 255:red, 74; green, 144; blue, 226 }  ,opacity=1 ]  {$\hat{\theta }_{t}$};
		\draw (328.1,218.2) node [anchor=north west][inner sep=0.75pt]  [font=\large,color={rgb, 255:red, 74; green, 144; blue, 226 }  ,opacity=1 ]  {$\Theta _{t},\delta _{\theta ,t}$};
		\draw (96.2,148) node [anchor=north west][inner sep=0.75pt]  [font=\normalsize, opacity = 1] [align=left] {\begin{minipage}[lt]{70.19pt}\setlength\topsep{0pt}
				\begin{center}
					\textcolor{black!60}{Nominal policy} 
				\end{center}
				
		\end{minipage}};
		\draw (31,160.4) node [anchor=north west][inner sep=0.75pt]  [font=\large,color={rgb, 255:red, 139; green, 87; blue, 42 }  ,opacity=1 ]  {$\mathcal{U}_{\text{ras} ,t}$};
		\draw (122,97) node [anchor=north west][inner sep=0.75pt]  [font=\normalsize] [align=left] {Safety filter \eqref{eq:sec3.2-safety_filter}};
		\draw (64,79.4) node [anchor=north west][inner sep=0.75pt]  [font=\large,color={rgb, 255:red, 208; green, 2; blue, 27 }  ,opacity=1 ]  {$u_{\text{nom} ,t}$};
		\draw (104,191) node [anchor=north west][inner sep=0.75pt]  [font=\normalsize] [align=left] {\begin{minipage}[lt]{92.31pt}\setlength\topsep{0pt}
				\begin{center}
					Input condition \eqref{eq:sec3.2-safe_control_condition_online} \\in Theorem 2
				\end{center}
				
		\end{minipage}};
		\draw (393,45) node [anchor=north west][inner sep=0.75pt]  [font=\large]  {$z^{-1}$};
		\draw (193,148.4) node [anchor=north west][inner sep=0.75pt]  [font=\large, opacity = 1]  {\textcolor{black!60}{$\pi _{\text{nom} ,t}$}};
		\draw  [draw opacity=0][fill={rgb, 255:red, 80; green, 227; blue, 194 }  ,fill opacity=1 ]  (33.2,50.4) -- (131.2,50.4) -- (131.2,70.4) -- (33.2,70.4) -- cycle  ;
		\draw (36.2,54.4) node [anchor=north west][inner sep=0.75pt]  [font=\normalsize] [align=left] {Safe controller};
		\draw (105,260) node [anchor=north west][inner sep=0.75pt]  [font=\normalsize,color={rgb, 255:red, 0; green, 0; blue, 0 }  ,opacity=1 ] [align=left] {\begin{minipage}[lt]{250.5pt}\setlength\topsep{0pt}
		\begin{center}
			\textcolor{black!60}{[Prerequisite Step] Design  of robust adaptive DT-CBF satisfying certificate~\eqref{eq:sec3.2-CBF_condition_ra} in Definition 4}
		\end{center}
				
		\end{minipage}};
		\draw (470,213.4) node [anchor=north west][inner sep=0.75pt]  [font=\large,color={rgb, 255:red, 74; green, 144; blue, 226 }  ,opacity=1 ]  {$\overline{\delta }_{\theta }$};
		\draw (46,273.4) node [anchor=north west][inner sep=0.75pt]  [font=\large]  {$\Theta $};
		\draw (489,259.4) node [anchor=north west][inner sep=0.75pt]  [font=\large,color={rgb, 255:red, 65; green, 117; blue, 5 }  ,opacity=1 ]  {$f$};
		\draw (174,231.4) node [anchor=north west][inner sep=0.75pt]  [font=\large,color={rgb, 255:red, 189; green, 16; blue, 224 }  ,opacity=1 ]  {$B,\alpha, \gamma$};
		\draw (46,255.4) node [anchor=north west][inner sep=0.75pt]  [font=\large]  {$\overline{w}$};
		\draw (347,93) node [anchor=north west][inner sep=0.75pt]   [align=left] {System};
		\draw (399,87) node [anchor=north west][inner sep=0.75pt]  [font=\large]  {$f\left( x,u;\theta ^{\ast }\right)$};

	\end{tikzpicture}

	}
	\label{fig:diagram_racbf}
	\vspace*{-0.4cm}
	\caption{Block diagram of the proposed robust adaptive safe control framework. Thick arrows (orange and blue) denote information exchange between the safe controller and the system/estimator ($\est$). The nominal policy and DT-CBF design blocks are rendered grayed out to indicate that they lie outside the scope of the current paper.}
\end{figure}

%% file: 3-main_content/2-preliminaries.tex
\vspace*{-0.5cm}
\section{Preliminaries}
\label{sec2:preliminaries}
\vspace*{-0.3cm}
This section establishes the foundation for the main results. We first introduce the system setting, then present robust DT-CBFs, showing how they define safe sets and enforce state safety through input feasibility via positive set invariance. In essence, CBFs constrain admissible controls to keep the state within the safe set or drive it back when it deviates. Finally, we introduce a class of online parameter estimators that provide (i) point estimates, (ii) set estimates, and (iii) uniform bounds on point-estimate increments.
\vspace*{-0.2cm}
\subsection{Notation}
\vspace*{-0.4cm}
Let $\bR$ and $\bN$ denote the sets of real and natural numbers, respectively, with $\obN := \bN \cup \{+\infty\}$ and $\bI_{[a:b]} := \obN \cap [a, b]$. The matrix or vector $p$-norm is $\|\cdot\|_p$, where $\|\cdot\|$ defaults to the $2$-norm. For a vector $x$, $x[i]$ is its $i$-th element. The zero vector, one vector, and identity matrix of size $n$ are denoted, respectively, by $\bzero_{n}$, $\bone_{n}$, and $\mathbf{I}_{n}$. A continuous function $\alpha:\bR \to \bR$ is an extended Class $\mathcal{K}$ function (i.e., $\alpha \in \ke$) if $\alpha(0) = 0$ and it is strictly increasing. The class of strictly sub-identity functions $\csi$ contains $\beta: \bR \to \bR$ such that $\beta(r) < r$ for all $r > 0$. For a closed set $\cX \subseteq \bR^n$, define $\Pi_{\cX}(\bar{x}) := \arg\min_{x \in \cX}\|x - \bar{x}\|$ and $\dist_{\cX}(\bar{x}) := \min_{x \in \cX}\|x - \bar{x}\|$, respectively. Finally, $\lambda_{\min}(\Gamma)$ denotes the minimum eigenvalue of a symmetric positive-definite matrix $\Gamma$.

\vspace*{-0.2cm}
\subsection{System description}
\label{sec2.1:system}
\vspace*{-0.4cm}
Consider a class of input-affine discrete-time systems:
\vspace*{-0.3cm}
\begin{align}
	\label{eq:sec2.2-model}
	x_{t+1} &= f(x_t,u_t;\para) + w_t \notag \\
	&= \fd(x_t) - [\bm{\phi}(x_t)]^\top \para + g(x_t)u_t + w_t,
\end{align}

\vspace*{-0.6cm}
\noindent where $x_t \in \cX \subset \bR^n$ and $u_t \in \cU \subseteq \bR^m$ are the state and input at time step $t$. The set $\cX$ is compact, and $\cU$ is closed to ensure the existence of solutions for the safety filter optimization problem \cite{ames2016control}, \cite{lopez2020robust}. These settings are practically valid for most engineering systems, where physical constraints or design specifications naturally confine the state and input variables to a bounded operational regime~\cite{agrawal2017discrete}, \cite{ames2016control}, \cite{lopez2020robust}. The model $\fd$ captures known dynamics~\cite{cosner2024bounding} or discretization effects~\cite{mania2022active}, while $g$ characterizes the input coupling. The term $[\bm{\phi}(x_t)]^\top\theta$ represents the uncertain drift, which models unknown damping or state-coupled disturbances \cite{luo2022sample}, \cite{taylor2020adaptive} through a known continuous kernel $\bm{\phi}$ and an unknown constant parameter $\para \in \Theta$. Accurate values for $\fd, \bm{\phi},$ and $g$ can be obtained via full state measurement. In addition, the disturbance $w_t$ lies in a known polytope\hspace*{-0.2em}\footnote{A polytope is a \textit{bounded} polyhedron (i.e., an intersection of a finite set of closed half-spaces)~\cite[Chapter 2.1]{borrelli2017predictive}.}given by
\vspace*{-0.2cm}
\begin{equation}
	\label{eq:sec2.2-disturbance_set}
	\cW = \{w \in \bR^n \mid H_w w \leq \bone_{n_w}\},
	\vspace*{-0.2cm}
\end{equation}
where $H_w \in \bR^{n_w \times n}$ and $n_w \geq n$ is the number of linear inequalities. It is clear that $\mathbf{0}_n$ lies in the interior of $\cW$, and we further define 
$\overline{w} := \max_{w \in \cW}\|w\|$. The true parameter $\tpara \in \setpara$ is a \textit{constant} and the prior known parameter set $\setpara$ is a polytope given by
\vspace*{-0.2cm}
\begin{equation}
	\label{eq:sec2.2-parameter_set}
	\setpara = \{\para \in \bR^q \mid H_{\para}\para \leq h_{\para}\},
	\vspace*{-0.2cm}
\end{equation}
where $H_{\para} \in \bR^{n_{\para}\times q}$ with $n_{\para} \geq q$, and $h_\para \in \bR^{n_{\para}}$. Both the disturbance and the model parameter are considered to lie in polytopes, which is motivated by practical considerations, as many disturbances and parameters are naturally bounded and can be modeled by box-constrained sets~\cite{ackermann1993robust}, \cite{ames2016control}, \cite{lopez2020robust}. Theoretically, polytopes can also inner-approximate arbitrary compact sets, making this a flexible and effective modeling choice. Given an estimate $\hat{\theta} \in \Theta$, its associated estimation error is defined as
\vspace*{-0.3cm}
\begin{equation}
	\label{eq:sec2.2-estimation_error}
	\epa := \hat{\theta} - \theta^\ast.
	\vspace*{-0.3cm}
\end{equation}
The estimation error $\epa$ cannot be computed exactly since $\tpara$ is unknown. However, for a specific $\hpara$, an upper bound on its associated estimation error can be computed by solving the following optimization problem:
\vspace*{-0.3cm}
\begin{equation}
	\label{eq:sec2.2-maximized_error}
	\be(p) := \max_{\theta \in \Theta} \|\theta - \hat{\theta}\|_p,
	\vspace*{-0.2cm}
\end{equation}
and it holds that\footnote{Given $x \in \bR^n$, $\|x\|_2 \leq \|x\|_1$ \cite{hornjohnson2012}.}$\|\epa\| \leq \be(2) \leq \be(1)$. Following \eqref{eq:sec2.2-model}, the \textit{true} system under disturbances is given by
\vspace*{-0.3cm}
\begin{equation}
	\label{eq:sec2.2-model_perturbed_true}
	x_{t+1} = f(x_t,u_t;\tpara) + w_t.
	\vspace*{-0.2cm}
\end{equation}
The system state evolves according to \eqref{eq:sec2.2-model_perturbed_true}; however, the controller does not have access to the true parameter~$\tpara$.

\begin{remark}
	The controller design for \eqref{eq:sec2.2-model} does not require smooth kernels; with non-smooth kernels, the model can represent PieceWise-Affine (PWA) dynamics~\cite{mania2022active}. The input-affine structure and linear parameter dependence are versatile, standard modeling choices~\cite{kohler2021robust}, \cite{nijmeijer1990nonlinear} that do not trivialize the safe control problem. This framework is widely used in application examples such as inverted pendulums \cite{luo2022sample} and vehicle cruise control \cite{taylor2020adaptive}.
\end{remark}

\vspace*{-0.2cm}
\subsection{Robust safe control}
\label{sec2.2:robust_safety}
\vspace*{-0.3cm}
This subsection introduces robust DT-CBFs, which constitute a natural and modest extension of continuous-time CBFs to discrete-time systems with bounded disturbances, building on prior work on continuous-time systems with bounded disturbances \cite{jankovic2018robust} and discrete-time systems without disturbances \cite{agrawal2017discrete}.
\vspace*{-0.2cm}
\begin{definition}[Robust safety]
	\label{def:invariance_safety_nominal}
	A set $\cXs \subseteq \cX$ is positive invariant for \eqref{eq:sec2.2-model_perturbed_true} if $x_0 \in \cXs$ implies that $\forall \{w_t\}^{\infty}_{t=0} \in \cW^{\infty}$, $x_{t} \in \cXs$ for all $t \in \bN$. In this case, the system \eqref{eq:sec2.2-model_perturbed_true} is robustly safe w.r.t.~$\cXs$, and $\cXs$ is called a robust safe set.
\end{definition}
\vspace*{-0.2cm}
Definition~\ref{def:invariance_safety_nominal} addresses only robustness against disturbances. It applies to the true parameter $\tpara$ but need not hold for all parameters $\para \in \setpara$. Characterizing safe sets using CBFs is useful for safe controller design~\cite{ames2016control}. For discrete-time systems, DT-CBFs are powerful tools to achieve this objective \cite{agrawal2017discrete}, \cite{cosner2024bounding}. Specifically, for a continuous barrier function $B: \cX \times \setpara \to \bR$, its induced safe set is defined using the $0$-superlevel set of $B(\cdot,\para):\cX \to \bR$ for all $\para \in \setpara$ as 
\vspace*{-0.2cm}
\begin{equation}
	\label{eq:sec2.3-safe_set}
	\cSb{\para} := \{x \in \cX \mid B(x,\para) \geq 0\},
	\vspace*{-0.2cm}
\end{equation}
and its boundary $\partial \cSb{\para}$ and interior $\inte(\cSb{\para})$ are defined, respectively, by $\partial \cSb{\para} := \{x \in \cX \mid B(x,\para) = 0\}$ and $\inte(\cSb{\para}) := \{x \in \cX \mid B(x,\para) > 0\}$.
\begin{remark}
	\label{rmk:barrier_function_dependency}
	This work adopts barrier functions that depend on the model parameter $\para$ as in~\cite{lopez2020robust} and~\cite{taylor2020adaptive} since most barrier functions are model-based~\cite{ames2016control}. Nonetheless, CBFs that are derived mainly from the safety requirements without using $\theta$ are also commonly used in safe control applications, e.g., adaptive cruise control~\cite{taylor2020adaptive} and aircraft pitch control~\cite{lopez2020robust}. 
\end{remark}
\vspace*{-0.2cm}
\begin{definition}[Lipschitz continuity]
	\label{ass:lipschitz_base_cbf}
	The function $B: \cX \times \setpara \to \bR$ is locally Lipschitz continuous on the joint space $\cX \times \setpara$ if there exist constants $L_{B,x}, L_{B,\para} \geq 0$ such that $\forall x',x'' \in \cX, \;\para',\para''\in\setpara$,
	\vspace*{-0.2cm}
	\begin{equation*}
		\label{eq:sec3.2-lipschitz}
		|B(x',\para') \sminus B(x'',\para'')| \leq L_{B,x}\|x' \sminus x''\| \splus L_{B,\para}\|\para' \sminus \para''\|.
	\end{equation*}
\end{definition}
\vspace*{-0.5cm}
\begin{definition}[Robust DT-CBF]
	\label{def:cbf_nominal}
	The function $B_{\tpara} := B(\cdot,\tpara)$ is a robust DT-CBF for \eqref{eq:sec2.2-model_perturbed_true} on $\cSb{\tpara}$ if it is Lipschitz continuous and there exists $\alpha \in \ke \cap \csi$ such that $\forall x \in \cX$, $\exists u \in \cU$ satisfying
	\vspace*{-0.2cm}
	\begin{equation}
		\label{eq:sec2.3-barrier_function}
		B_{\tpara}(f(x,u;\tpara)) \sminus B_{\tpara}(x) \sminus L_{B,x}\overline{w}\geq -\alpha\left(B_{\tpara}(x)\right),
		\vspace*{-0.2cm}
	\end{equation}
	and \eqref{eq:sec2.3-barrier_function} is called the robust control barrier certificate against disturbances.
\end{definition}
\vspace*{-0.2cm}
In discrete time, Lipschitz continuity is adopted to quantify the perturbation of $B$. The barrier certificate \eqref{eq:sec2.3-barrier_function} also defines the set of robust safe control inputs for $\cSb{\tpara}$ as
\vspace*{-0.2cm}
\begin{equation*}
	\label{eq:sec2.3-safe_input}
	\Usf(x) = \{u \in \cU \mid \eqref{eq:sec2.3-barrier_function} \text{ holds}\},
	\vspace*{-0.2cm}
\end{equation*}
where the subscript $\tpara$ emphasizes that characterizing $\Usf(x)$ requires knowing the true parameter. The following theorem is a standard result on CBF-based safety guarantees, and the proof is omitted for brevity since it follows the same procedure as in \cite{agrawal2017discrete}.
\vspace*{-0.2cm}
\begin{thm}[Robust safety]
	\label{thm:cbf_nominal}
	If the continuous function $B_{\tpara}: \cX \to \bR$ is a robust DT-CBF for \eqref{eq:sec2.2-model_perturbed_true} on $\cSb{\tpara}$, then $\Usf(x)$ is non-empty for all $x \in \cX$. In addition, if a state-feedback control policy $\pi: \cX \to \bR^m$ satisfies $\pi(x) \in \Usf(x)$, $\forall x \in \cX$, then the closed-loop system $x_{t+1} = f(x_t,\pi(x_t);\tpara) + w_t$ is robustly safe w.r.t.~$\cSb{\tpara}$ against disturbances $w_t$.
\end{thm}
\vspace*{-0.2cm}
Since knowing $\tpara$ is often unrealistic, the certificate \eqref{eq:sec2.3-barrier_function} cannot be directly used for safe control when $\tpara$ is unknown. A common remedy is to robustify against worst-case model mismatch by considering a constant nominal $\hpara$ and adding terms that scale with the estimation error, yielding a worst-case safety guarantee similar to Theorem~\ref{thm:cbf_nominal}. However, such designs lead to conservative inputs and may have worse performance. Because parametric uncertainty is \textit{epistemic} and can be reduced using data~\cite{wang2025safe}, incorporating adaptive safety mechanisms with online parameter estimation is more desirable.

\vspace*{-0.3cm}
\subsection{Online parameter estimation}
\label{subsec:2.3-online_adaptation}
\vspace*{-0.3cm}
Given $\setpara$ in~\eqref{eq:sec2.2-parameter_set}, an estimator $\est$ provides a point estimate $\hpa{t}$ and a polytopic set $\setpara_t \subseteq \setpara$ at each time step $t$ such that $\tpara \in \setpara_t$. These sets are initialized as $\setpara_0 = \setpara$ and recursively refined, for instance, via set-membership identification. Leveraging \eqref{eq:sec2.2-estimation_error} and \eqref{eq:sec2.2-maximized_error}, the instantaneous estimation error and its associated bound are defined as
\vspace*{-0.3cm}
\begin{subequations}
	\label{eq:sec3.1-time_varying_update}
	\begin{align}
		\label{eq:sec3.1-time_varying_update_error}
		& \epat{t} := \hpa{t} - \tpara, \\
		\label{eq:sec3.1-time_varying_update_bound}
		& \bet{t}(p) := \max_{\para \in \setpara_t}\|\para - \hpa{t}\|_{p}.
	\end{align}
\end{subequations}

\vspace{-0.8cm}
\noindent The module $\est$ performs recursive updates according to the following abstracted law:
\vspace*{-0.4cm}
\begin{equation}
	\label{eq:sec3.1-general_update_law}
	\hpa{t+1} = \cT_{\adp,t}(\hpa{t}),
\end{equation}

\vspace{-0.7cm}
\noindent where the time-varying operator $\cT_{\adp,t}$ depends on online closed-loop data and the set estimate $\setpara_t$. The estimate increment is defined as
\vspace*{-0.3cm}
\begin{equation}
	\label{eq:sec3.1-estimate_increment}
	\ipa{t} := \hpa{t+1} - \hpa{t} = \cT_{\adp,t}(\hpa{t}) - \hpa{t}.
\end{equation}
\vspace{-1cm}
\begin{assumption}
	\label{ass:bounded_increments_assumption}
	There exists a uniform upper bound $\bar{\delta}_{\para}$ on the estimate increments $\ipa{t}$, i.e., $\|\ipa{t}\| \leq \bar{\delta}_{\para}, \forall t \in \obN$.
\end{assumption}
\vspace{-0.3cm}
\noindent Since $\Theta$ is compact, Assumption~\ref{ass:bounded_increments_assumption} holds trivially since $\|\ipa{t}\| \leq \max_{\theta', \theta'' \in \Theta} \|\theta' - \theta''\|$, i.e., the diameter of $\Theta$.

%% file: 3-main_content/3-adaptive_cbf.tex
\vspace*{-0.2cm}
\section{Robust adaptive safe control}
\vspace*{-0.3cm}
\label{sec3:robust_adaptive_safety}
This section introduces robust adaptive DT-CBFs, providing a sufficient condition for their certification. Building on this, we derive an online input condition to compute safe controls within a safety filter framework. Finally, we discuss some practical implementation issues.


\input{3-main_content/3.1-problem_formulation}
\input{3-main_content/3.2-adaptive_cbf_core}
\input{3-main_content/3.3-practical_implementation}

%% file: 3-main_content/3.1-problem_formulation.tex
\vspace*{-0.2cm}
\subsection{Problem formulation}
\label{subsec:3.1-problem_formulation}
\vspace*{-0.3cm}
Given a candidate function $B: \mathcal{X} \times \Theta \to \mathbb{R}$ defining the safe set in \eqref{eq:sec2.3-safe_set}, we first provide the conditions required for $B$ to qualify as a \textit{robust adaptive DT-CBF}.
\begin{definition}[Robust adaptive DT-CBF]
	\label{def:cbf_ra}
	Given an estimator $\est$ satisfying Assumption~\ref{ass:bounded_increments_assumption}, a barrier function $B: \cX\times \setpara \to \bR$ is a robust adaptive DT-CBF for \eqref{eq:sec2.2-model_perturbed_true} and $\est$ if it is Lipschitz continuous and there exists $\alpha \in \ke \cap \mnt{\csi}$ such that, for all $x \in \cX$, there exists $u \in \cU$ satisfying
	\vspace*{-0.3cm}
	\begin{multline}
		\label{eq:sec3.2-CBF_condition_ra}
		\Delta B{^\star}(x,u) - L_{B,x}\overline{w} - E^{\star}_{\para}(x) \\ \geq -\alpha\left(B^{\star}(x) - \frac{(\bar{\varepsilon}_{\theta}(p))^2}{2\gamma} \right),
	\end{multline}
	
	\vspace{-0.6cm}
	\noindent where $\gamma > 0$, $B^{\star}(x) := \min_{\para \in \setpara}B(x,\para)$, $\Delta B{^\star}(x,u) \sdeq \min_{\para\in \Theta}\{B(f(x,u;\para),\para) - B(x,\para)\}$, $\bar{\varepsilon}_{\theta}(p)$ is the worst-case error $p$-norm ($p \in \{1,2\}$) defined as
	\vspace*{-0.2cm}
	\begin{equation}
		\label{eq:sec3.2-def_cbc_error_bound}
		\bar{\varepsilon}_{\theta}(p) := \max_{\theta,\theta' \in \Theta}\|\theta - \theta'\|_p,
		\vspace*{-0.2cm}
	\end{equation}
	
	\vspace*{-0.6cm} 
	\noindent and $E^{\star}_{\theta}(x) = \left(L_{B,x}\|\bm{\phi}(x)\| + \frac{\bar{\delta}_{\theta}}{\gamma}\right)\bar{\varepsilon}_{\theta}(p) + \frac{\bar{\delta}_{\theta}^2}{2\gamma} + L_{B,\para}\bar{\delta}_{\theta}$, with $L_{B,x}$ and $L_{B,\para}$ given in Definition~\ref{ass:lipschitz_base_cbf} and $\bar{\delta}_{\para}$ characterized by $\est$ as in Assumption~\ref{ass:bounded_increments_assumption}. 
	In this case, the condition \eqref{eq:sec3.2-CBF_condition_ra} is called the robust adaptive Control Barrier Certificate (CBC) against disturbances and parametric model mismatch.
\end{definition}
\vspace*{-0.2cm}
\begin{assumption} \label{asm:CBF_existence}
	For the model~\eqref{eq:sec2.2-model} with sets $\cW$ and $\Theta$ defined in~\eqref{eq:sec2.2-disturbance_set} and~\eqref{eq:sec2.2-parameter_set}, respectively, and an estimator $\est$ satisfying Assumption~\ref{ass:bounded_increments_assumption}, there exists a known robust adaptive DT-CBF $B: \cX \times \setpara \to \bR$ available for controller synthesis.
\end{assumption}
\vspace*{-0.3cm}
The condition \eqref{eq:sec3.2-CBF_condition_ra} provides a discrete-time certificate without using $\theta^*$, extending the framework in~\cite{lopez2020robust}. Assumption~\ref{asm:CBF_existence} is a standard feasibility condition~\cite{agrawal2017discrete}, \cite{lopez2020robust}, ensuring that a safety certificate exists and that the underlying optimization problem is well-posed. In practice, such functions are typically obtained by parameterizing $B$ and solving for its coefficients via numerical optimization~\cite{agrawal2017discrete}, \cite{ames2016control}. Assuming a valid robust adaptive DT-CBF is available, the objective is (i) to design a less conservative condition to certify safe inputs with parameter estimation and (ii) to use this condition to modify a potentially unsafe nominal policy $\pi_{\mathrm{nom},t}: \mathcal{X} \to \mathbb{R}^m$ such that the constraints encoded by the DT-CBF are satisfied. 
\vspace*{-0.2cm}
\begin{remark}
	Although the Lipschitz-continuity requirement restricts the search space, it is standard in the literature because commonly used CBFs are continuously differentiable~\cite{lopez2020robust}, \cite{taylor2020adaptive}, implying the required Lipschitz continuity. In practice, this assumption is not restrictive; in fact, Lipschitz-continuous DT-CBFs are widely used in several discrete-time safe-control applications \cite{agrawal2017discrete}, \cite{cosner2024bounding}, \cite{luo2022sample}.
\end{remark}
\vspace*{-0.2cm}
\begin{remark}
	The considered nominal controller may either ignore safety entirely during design (e.g., basic PID controllers) or incorporate safety constraints without formal guarantees. Examples of the latter include model-based RL with safety-based rewards or approximate MPC that does not inherently preserve the safety properties of the original MPC controller~\cite{hertneck2018learning}.
\end{remark}

%% file: 3-main_content/3.2-adaptive_cbf_core.tex
\vspace*{-0.1cm}
\subsection{Robust adaptive safe control}
\label{subsec:3.2-robust_barrier_certificate}
\vspace*{-0.3cm}
To address parametric uncertainty, we define the \textit{robust adaptive} counterpart of a given $B$ with estimate $\hpa{t}$ by
\vspace*{-0.3cm}
\begin{equation}
	\label{eq:sec3.2-ra_barrier_function}
	\Brt{t}(x) := B(x,\hpara_t) - \frac{1}{2}\epat{t}^\top\Gamma^{-1}\epat{t},
	\vspace*{-0.2cm}
\end{equation}
where $\Gamma$ is a symmetric positive definite matrix such that $\mG = \gamma$ with $\gamma$ given in~\eqref{eq:sec3.2-CBF_condition_ra}. Accordingly, the \textit{robust adaptive} safe set $\cSr{t}$ is defined using $\Brt{t}(\cdot)$ as
\vspace*{-0.2cm}
\begin{equation}
	\label{eq:sec3.2-ra_safe_set}
	\cSr{t} := \{x \in \cX \mid \Brt{t}(x) \geq 0\}.
	\vspace*{-0.2cm}
\end{equation}
It is noted that $\cSr{t} \subseteq \cSb{\hpara_t}$ (see Fig.~\ref{fig:subA}), and $\cSr{t} = \cSb{\tpara}$ if and only if $\hpa{t} = \tpara$. 
\vspace*{-0.2cm}
\begin{definition}[Timed robust safety]
	\label{def:invariance_safety_time}
	Given $T \in \obN$, $\{\cXst{t}\}^{T}_{t=0}$ satisfying $x_0 \in \cXst{0}$ and $\forall t \in \bI_{[0:T]}$, $\cXst{t} \subseteq \cX$ is sequentially positively invariant for \eqref{eq:sec2.2-model_perturbed_true} if $\forall t \in \bI_{[0:T-1]}$, $x_{t} \in \cXst{t}$ implies that $x_{t+1} \in \cXst{t+1}$  for all $w_t \in \cW$. In this case, the system \eqref{eq:sec2.2-model_perturbed_true} is said to be robustly safe w.r.t.~$\{\cXst{t}\}^{T}_{t=0}$. In addition, if $\cXst{t} \subseteq \cXs^\ast$, $\forall t \in \bI_{[0:T]}$, the system \eqref{eq:sec2.2-model_perturbed_true} is said to be robustly safe w.r.t.~$\cXs^\ast$ over $\bI_{[0:T]}$.
\end{definition}
\input{5-figures/fig_sketches}
Finite-horizon safety has been widely studied in CBF-based control \cite{cosner2024bounding}. The timed safety notion in Definition~\ref{def:invariance_safety_time} is both more practical and more general than its infinite-horizon counterpart, as safety is typically required only over a finite operational period \cite{santoyo2019verification}; setting $T=+\infty$ recovers infinite-horizon safety.

\vspace*{-0.2cm}
\begin{thm}[Robust adaptive safety]
	\label{thm:cbf_ra}
	Given $x_0 \in \cSr{0}$, a horizon $T \in \obN$, an estimator $\est$ satisfying Assumption~\ref{ass:bounded_increments_assumption}, and a robust adaptive DT-CBF $B: \cX\times \setpara \to \bR$ with $\gamma > 0$ and function $\alpha \in \ke \cap \mnt{\csi}$, for the robust adaptive online input condition
	\vspace*{-0.4cm}
	\begin{multline}
		\label{eq:sec3.2-safe_control_condition_online}
		B\big(f(x,u;\hpa{t}),\hpa{t}\big) - B(x,\hpa{t}) - L_{B,x}\overline{w} \\ -E_{\theta,t}(x)  \geq -\alpha\left(B(x,\hpara_t) - \frac{(\bet{t}(p))^2}{2\gamma}\right),
	\end{multline}
	
	\vspace{-0.7cm}
	\noindent the admissible input set $\Uasf{t}(x;p)$ defined as
	\vspace*{-0.2cm}
	\begin{equation}
		\label{eq:sec3.2-safe_control_set_online}
		\Uasf{t}(x;p) := \left\{u \in \cU \mid \eqref{eq:sec3.2-safe_control_condition_online} \text{ holds}\right\}
		\vspace*{-0.2cm}
	\end{equation}
	is non-empty, $\forall x \in \cX$ and $t \in \bI_{[0,T-1]}$. The term $E_{\para,t}(x) = \left(L_{B,x}\|\bm{\phi}(x)\| + \frac{\|\ipa{t}\|}{\gamma}\right)\bet{t}(p) +L_{B,\theta}\|\ipa{t}\|+ \frac{\|\ipa{t}\|^2}{2\gamma}$ with $\ipa{t}$ defined in \eqref{eq:sec3.1-estimate_increment}. In addition, if a time-varying control policy $\pi_t: \cX \to \bR^m$ satisfies $\pi_t(x) \in \Uasf{t}(x;p)$ for all $x \in \cX$ and $t \in \bI_{[0,T-1]}$, then the time-varying closed-loop system
	\vspace*{-0.2cm}
	\begin{equation}
		\label{eq:sec3.2-closed_theorem_example}
		x_{t+1} = f(x_t,\pi_t(x_t);\tpara) + w_t
		\vspace*{-0.2cm}
	\end{equation}
	is robustly safe w.r.t.~$\{\cSr{t}\}^{T}_{t=0}$ and thus robustly safe w.r.t.~$\{\cSb{\hpara_t}\}^{T}_{t=0}$ (cf. Definition \ref{def:invariance_safety_time}) against both parametric model mismatch and disturbances.
\end{thm}
\vspace{-0.2cm}
The proof of Theorem~\ref{thm:cbf_ra} is provided in Appendix~\ref{app:proof_1_theorem_2}. Notably, condition~\eqref{eq:sec3.2-CBF_condition_ra} guarantees that the admissible set $\Uasf{t}(x;p)$ is non-empty, while the online safe control condition~\eqref{eq:sec3.2-safe_control_condition_online} depends solely on estimated parameters. To ensure safety with minimal intervention, we employ a safety filter that projects the nominal input onto the admissible set (see Fig.~\ref{fig:subB}). Specifically, for a nominal policy $\pi_{\mathrm{nom},t}: \cX \to \bR^m$, the robust adaptive safe policy $\uasf{t}: \cX \to \mathcal{U}$ is implicitly defined by the following constrained optimization:
\vspace*{-0.2cm}
\begin{equation}
	\label{eq:sec3.2-safety_filter}
	\uasf{t}(x;p) = \argmin_{\nu \in \Uasf{t}(x;p)}\|\nu - \pi_{\mathrm{nom},t}(x)\|.
	\vspace*{-0.2cm}
\end{equation}
This policy is time-varying as $\Uasf{t}$ evolves to capture historical state and input data via the estimator $\est$.

\vspace*{-0.3cm}
\begin{remark}
	The safe initial state condition ($x_0 \in \cSr{0}$), verified via the error bound $\bet{0}(p)$, does not restrict the argument of $\alpha(\cdot)$ in~\eqref{eq:sec3.2-CBF_condition_ra} and~\eqref{eq:sec3.2-safe_control_condition_online} to positive values. For instance, condition~\eqref{eq:sec3.2-CBF_condition_ra} remains valid whenever a feasible input $u \in \mathcal{U}$ yields a sufficiently large $\Delta B^\star(x,u)$, regardless of the sign of the argument of $\alpha(\cdot)$.
\end{remark}
\vspace*{-0.3cm}
\begin{remark}
	\label{rmk:error_choice}
	The online input condition \eqref{eq:sec3.2-safe_control_condition_online} is less conservative than the CBC \eqref{eq:sec3.2-CBF_condition_ra}, which is reserved for offline verification. In contrast to \cite{lopez2020robust}, the robustified function \eqref{eq:sec3.2-ra_barrier_function} employs the true error $\epat{t}$ instead of a set-induced bound on $\epat{t}$. Notably, the invariance property for robustified CBFs using bounds on $\epat{t}$ is valid in discrete time but not in continuous time; see Appendix \ref{app:subtlety}.
\end{remark}

%% file: 5-figures/fig_sketches.tex
\begin{figure}[h]
	\centering
	\tikzset{every picture/.style={line width=0.75pt}} 
	\begin{subfigure}[b]{0.4\columnwidth}
		\centering
		\resizebox{0.75\columnwidth}{!}{%
		
		\begin{tikzpicture}[x=0.75pt,y=0.75pt,yscale=-1,xscale=1]
			
			\draw  [fill={rgb, 255:red, 155; green, 155; blue, 155 }  ,fill opacity=0.11 ] (204.6,45.12) -- (259.2,58.49) -- (279.6,116.12) -- (141.6,104.09) -- cycle ;
			\draw  [fill={rgb, 255:red, 248; green, 231; blue, 28 }  ,fill opacity=0.16 ] (195.2,60.6) .. controls (215.2,50.6) and (225.2,48.2) .. (249.6,62.2) .. controls (274,76.2) and (239.2,111) .. (216.8,108.6) .. controls (194.4,106.2) and (180,99.8) .. (175.2,91) .. controls (170.4,82.2) and (175.2,70.6) .. (195.2,60.6) -- cycle ;
			\draw  [fill={rgb, 255:red, 208; green, 2; blue, 27 }  ,fill opacity=0.11 ] (194.4,68.09) .. controls (212,58.49) and (225.6,54.49) .. (244.8,66.49) .. controls (264,78.49) and (232,104.02) .. (217.6,102.49) .. controls (203.2,100.96) and (193.6,98.56) .. (186.4,92.89) .. controls (179.2,87.22) and (176.8,77.69) .. (194.4,68.09) -- cycle ;
			\draw [color={rgb, 255:red, 152; green, 143; blue, 33 }  ,draw opacity=1 ] [dash pattern={on 0.84pt off 2.51pt}]  (178.91,75.95) .. controls (167.86,75.43) and (164.88,67.07) .. (162.82,61.96) ;
			\draw [shift={(161.6,59.32)}, rotate = 59.04] [fill={rgb, 255:red, 152; green, 143; blue, 33 }  ,fill opacity=1 ][line width=0.08]  [draw opacity=0] (8.04,-3.86) -- (0,0) -- (8.04,3.86) -- (5.34,0) -- cycle    ;
			
			\draw (153.4,89.8) node [anchor=north west][inner sep=0.75pt]  [font=\large]  {$\mathcal{X}$};
			\draw (209,66.6) node [anchor=north west][inner sep=0.75pt]  [font=\small,color={rgb, 255:red, 208; green, 2; blue, 27 }  ,opacity=0.51 ]  [font=\large] {$\mathcal{S}_{\text{ra} ,t}$};
			\draw (121,44) node [anchor=north west][inner sep=0.75pt]  [color={rgb, 255:red, 191; green, 178; blue, 35 }  ,opacity=1 ] [font=\large] {$\mathcal{S}_{B,\hat{\theta }_{t}}$};

		\end{tikzpicture}
		}
		\caption{Robustification}
		\label{fig:subA}
	\end{subfigure}
	\hfill
	\begin{subfigure}[b]{0.4\columnwidth}
		\centering
		\resizebox{0.75\columnwidth}{!}{%

		\tikzset{every picture/.style={line width=0.75pt}} 
		
		\begin{tikzpicture}[x=0.75pt,y=0.75pt,yscale=-1,xscale=1]
			
			\draw  [fill={rgb, 255:red, 189; green, 16; blue, 224 }  ,fill opacity=0.07 ] (123,35) -- (217.32,38.82) -- (218.32,107.82) -- (75.32,77.82) -- cycle ;
			\draw  [dash pattern={on 4.5pt off 4.5pt}]  (113.32,56.82) -- (165.32,32.82) ;
			\draw    (126.71,15.29) -- (138.17,40.34) ;
			\draw [shift={(139,42.15)}, rotate = 245.41] [fill={rgb, 255:red, 0; green, 0; blue, 0 }  ][line width=0.08]  [draw opacity=0] (8.4,-2.1) -- (0,0) -- (8.4,2.1) -- cycle    ;
			\draw  [fill={rgb, 255:red, 74; green, 144; blue, 226 }  ,fill opacity=1 ] (123.15,14) .. controls (123.15,12.43) and (124.43,11.15) .. (126,11.15) .. controls (127.57,11.15) and (128.85,12.43) .. (128.85,14) .. controls (128.85,15.57) and (127.57,16.85) .. (126,16.85) .. controls (124.43,16.85) and (123.15,15.57) .. (123.15,14) -- cycle ;
			\draw    (138,39) -- (144.18,35.82) ;
			\draw    (144.18,35.82) -- (147.18,41.82) ;
			\draw  [color={rgb, 255:red, 65; green, 117; blue, 5 }  ,draw opacity=1 ][fill={rgb, 255:red, 184; green, 233; blue, 134 }  ,fill opacity=0.51 ] (140.15,45) .. controls (160.15,35) and (201.34,46.82) .. (207.34,54.82) .. controls (213.34,62.82) and (218.34,71.82) .. (209.34,83.82) .. controls (200.34,95.82) and (163.34,94.82) .. (136.34,81.82) .. controls (109.34,68.82) and (120.15,55) .. (140.15,45) -- cycle ;
			\draw  [fill={rgb, 255:red, 245; green, 166; blue, 35 }  ,fill opacity=1 ] (137.15,45) .. controls (137.15,43.43) and (138.43,42.15) .. (140,42.15) .. controls (141.57,42.15) and (142.85,43.43) .. (142.85,45) .. controls (142.85,46.57) and (141.57,47.85) .. (140,47.85) .. controls (138.43,47.85) and (137.15,46.57) .. (137.15,45) -- cycle ;
			
			\draw (145,2) node [anchor=north west][inner sep=0.75pt]  [font=\small,color={rgb, 255:red, 74; green, 144; blue, 226 }  ,opacity=1 ] [font=\large] {$\pi _{\text{nom} ,t}( x)$};
			\draw (39.15,27.4) node [anchor=north west][inner sep=0.75pt]  [font=\small,color={rgb, 255:red, 245; green, 166; blue, 35 }  ,opacity=1 ] [font=\large] {$\pi _{\text{ras} ,t}( x;p)$};
			\draw (127.6,59) node [anchor=north west][inner sep=0.75pt]  [font=\small,color={rgb, 255:red, 65; green, 117; blue, 5 }  ,opacity=1 ] [font=\large] {$\mathcal{U}_{\text{ras} ,t}( x;p)$};
			\draw (92,63.4) node [anchor=north west][inner sep=0.75pt] [font=\large] {$\mathcal{U}$};

		\end{tikzpicture}
		
		}
		\caption{Safety filter}
		\label{fig:subB}
	\end{subfigure}
	\vspace*{-0.2cm}
	\caption{Sketch of (a) adaptively robustified safe sets, and (b) the safety filter projection when the nominal input is unsafe.}
	\label{fig:twoTikz}
\end{figure}

%% file: 3-main_content/3.3-practical_implementation.tex
\vspace*{-0.2cm}
\subsection{Practical implementation}
\label{subsec:3.3-practical_implementation}
\vspace*{-0.3cm}
Deriving a closed-form policy $\uasf{t}(\cdot;p)$ is generally intractable. In practice, the safe control input $\uasf{t}(x_t;p)$ is computed by solving \eqref{eq:sec3.2-safety_filter} for the current state $x_t$. This requires characterizing $\Uasf{t}(x_t;p)$ via condition \eqref{eq:sec3.2-safe_control_condition_online}, which depends on $\bet{t}(p)$. While computing $\bet{t}(2)$ typically involves a non-convex program, using $\bet{t}(1)$ reduces the calculation to an efficient linear program. Since $\bet{t}(2) \leq \bet{t}(1)$, the $1$-norm-based condition is more conservative but computationally advantageous.

\vspace*{-0.2cm}
While \eqref{eq:sec3.2-safety_filter} is generally non-convex, the input-affine dynamics ensure $f$ is linear in $u$. If the barrier candidate $B$ is concave in $x$, then $\Uasf{t}(x;p)$ is a convex set, and computing $\uasf{t}(x_t;p)$ becomes a convex optimization problem with formal convergence and optimality guarantees \cite{boyd2004convex}. Although state-concavity is standard in DT-CBF research \cite{agrawal2017discrete}, \cite{zeng2021enhancing}, designing such functions is application-specific and remains outside the scope of this work.

\vspace*{-0.2cm}
\begin{remark}
	\label{rmk:norm}
	Employing $\bet{t}(1)$ in \eqref{eq:sec3.2-safe_control_condition_online} introduces conservatism, which increases the suboptimality of $\iasf{t}(1)$ compared to $\iasf{t}(2)$. This is particularly relevant when the nominal control policy is derived from optimal control frameworks such as discrete-time CLFs \cite{bof2018lyapunov}, model-based RL \cite{cohen2023safe}, or approximate MPC \cite{hertneck2018learning}, \cite{liu2025approximate}. An alternative approach involves uniformly adopting the $\infty$-norm for all error bounds, including equations \eqref{eq:sec2.2-maximized_error}, \eqref{eq:sec3.1-time_varying_update_bound}, and \eqref{eq:sec3.2-def_cbc_error_bound}, as well as the bounds in Definition~\ref{ass:lipschitz_base_cbf} and Assumption~\ref{ass:bounded_increments_assumption}.
\end{remark}

%% file: 3-main_content/4-online_rls.tex
\vspace*{-0.2cm}
\section{Online estimation and causal synthesis}
\vspace*{-0.3cm}
\label{sec4:identification}
Following the parameter estimator in \cite{lorenzen2019robust}, we instantiate the estimator $\est$ using a clipped recursive-least-squares (RLS) update with set-membership projection~\cite{milanese2013bounding}. We further present a \textit{causal} integration of the safety filter and estimator, ensuring that the computation of $\iasf{t}(p)$ relies solely on information available up to time step $t$.

\vspace*{-0.3cm}
For $t \geq 1$, both $\hpa{t}$ and $\setpara_t$ are determined by the information flow $\cF_{t-1} := (x_0, u_0, x_1, u_1, \dots, x_{t-1})$, and a causal synthesis means computing $u_t$ based on $\cF_t$. The initial estimate satisfies $\hpa{0} \in \setpara_0$ with $\setpara_0 = \setpara$. Given $x_t$, the residual is defined by $r_t := x_t - \fd(x_{t-1}) - g(x_{t-1})u_{t-1}$, and it holds that $r_t + [\bm{\phi}(x_{t-1})]^\top \theta^\ast \in \cW$, i.e., $H_{w}(r_t + [\bm{\phi}(x_{t-1})]^\top \theta^\ast) \leq \bone_{n_w}$ by \eqref{eq:sec2.2-disturbance_set}. Equivalently, $\tpara \in \Delta_{\para,t}$ with the non-falsified polytopic set $\Delta_{\theta,t}$ given by
\vspace*{-0.2cm}
\begin{equation}
	\label{eq:sec4-non_falsified_set}
    \Delta_{\theta,t} = \left\{\theta \in \bR^q \mid H_w[\bm{\phi}(x_{t-1})]^\top \theta \leq \bone_{n_w} \sminus H_w r_t\right\}.
    \vspace*{-0.2cm}
\end{equation}
Given the current set estimate $\setpara_t = \{\para \in \bR^q\mid H_{\para,t}\para \leq h_{\para,t}\}$, the updated set estimate is computed as
\vspace*{-0.2cm}
\begin{equation}
	\label{eq:sec4-set_update}
	\setpara_{t+1} = 
	\setpara_t \cap \Delta_{\theta,t} 
	\vspace*{-0.2cm}
\end{equation}
which is a polytope described by an augmented set of hyperplanes\footnote{Unbounded growth in the number of hyperplanes can be avoided by overapproximating the intersection with a hyperrectangle (or hyperbox) of fixed dimension~\cite{lorenzen2019robust}.}inherited from $\setpara_t$ and $\Delta_{\theta,t}$. For the point estimate, a \textit{prior} is first computed based on the clipped RLS parameter update as
\vspace*{-0.2cm}
\begin{equation}
	\label{eq:sec4-prior_update}
	\hpa{t+1}^{[\mathrm{p}]} = \hpa{t} - \mu_t \underbrace{\bm{\phi}(x_{t-1})\left(x_t - f(x_{t-1},u_{t-1};\hpa{t})\right)}_{:=\psi_{\para,t}},
	\vspace*{-0.2cm}
\end{equation}
where $\mu_t = \min\left\{(\left\|\psi_{\para,t}\right\| + \epsilon)^{-1}, \|\bm{\phi}(x_{t-1})\|^{-2}\right\}$ is the clipped learning rate with $\epsilon > 0$ being a small constant. Note that \eqref{eq:sec4-prior_update} recovers a gradient descent update \cite{lorenzen2019robust}, while the proposed clipped RLS update aligns with gradient clipping techniques~\cite{kingma2014adam}. The final estimate $\hpa{t+1}$ is obtained via a simple point-to-set projection:
\vspace*{-0.3cm}
\begin{equation}
	\label{eq:sec4-sm_projection}
	\hpa{t+1} = \Pi_{\setpara_{t+1}}(\hpa{t+1}^{[\mathrm{p}]}).
\end{equation}

\vspace*{-0.7cm}
\noindent Due to $\|\mu_t\psi_{\para,t}\| \leq 1$ and the triangle inequality, $\|\delta_{\para,t}\| \leq \min\{1, \max_{\theta', \theta'' \in \Theta} \|\theta' - \theta''\|\}$, consistent with Assumption~\ref{ass:bounded_increments_assumption}. As $u_t$ depends on $\mathcal{F}_t$, causality is ensured by aligning $\hat{\theta}_t$ and $\Theta_t$ with the delayed information flow $\mathcal{F}_{t-1}$.
\vspace*{-0.4cm}

Algorithm~\ref{alg:racbf} details the core procedure. At each step $t$, $\hpa{t}$ and $\ipa{t}$ are used to define $\Uasf{t}(x_t;p)$. A safe input $u_t$ is then computed via the safety filter, and the system state evolves to $x_{t+1}$. A summary of the proposed safe control approach is provided in Fig.~1. Note that the nominal controller may also use the estimates $\hpa{t}$ via certainty-equivalence model-based approaches~\cite{liu2024stability}, \cite{liu2025regret}.
\vspace*{-0.3cm}
\begin{remark}
	The adaptation bound in Assumption~\ref{ass:bounded_increments_assumption} is reminiscent of the averaging methods in classical adaptive control~\cite{astrom1994adaptive}. However, its role in the proposed method is not to perform averaging, but to directly limit the adaptation rate, which is necessary to prevent excessive growth of $B$ and thus to maintain the input condition~\eqref{eq:sec3.2-safe_control_condition_online}.
\end{remark}
\begin{remark}
	Convergence of $\hat{\theta}_t$ to $\tpara$ typically requires persistent excitation \cite{ljung2010perspectives}, often achieved through active exploration \cite{mania2022active}. In contrast, the proposed approach ensures safety by robustifying against the uncertainty set $\Theta_t$ containing $\tpara$. While convergence minimizes conservatism as $E_{\theta,t}(x)$ in \eqref{eq:sec3.2-safe_control_condition_online} approaches zero, our safety guarantees do not rely on achieving it. This work aims to ensure safety for a broad class of estimators; designing specific estimators or ensuring persistent excitation is beyond the scope of the current paper.
\end{remark}
\begin{algorithm}[h]
	\caption{Robust Adaptive Safe Control}
	\label{alg:racbf}
	\begin{algorithmic}
		\Require $B$, $\alpha$, $x_0$, $\hpa{0}$, $\setpara$, $\overline{w}$, $L_{B,x}$, $L_{B,\para}$, $\Gamma$
		\State $\setpara_0 \leftarrow \setpara$, $\mu_0 \leftarrow 0$, $\Delta_{\para,0} \leftarrow \bR^q$
		\For{$t = 0$ to $T-1$}
		\State Obtain nominal policy $\pi_{\mathrm{nom},t}$
		\State Compute $\bet{t}(p)$ \Comment{\eqref{eq:sec3.1-time_varying_update_bound}}
		\State Compute $\hpa{t+1}$ and $\delta_{\para,t}$ \Comment{\eqref{eq:sec4-non_falsified_set}--\eqref{eq:sec4-sm_projection}, \eqref{eq:sec3.1-estimate_increment}}
		\State Determine $\Uasf{t}(x;p)$  \Comment{\eqref{eq:sec3.2-safe_control_condition_online}--\eqref{eq:sec3.2-safe_control_set_online}}
		\State Compute $u_t$ using the safety filter \Comment{\eqref{eq:sec3.2-safety_filter}}
		\State Obtain $x_{t+1}$ by applying $u_t$
		\EndFor
	\end{algorithmic}
\end{algorithm}

%% file: 3-main_content/5-robustness.tex
\vspace*{-0.2cm}
\section{Inherent robustness}
\label{sec5:robustness}
\vspace*{-0.4cm}
Following the analysis in \cite{xu2015robustness}, this section examines the inherent robustness of \eqref{eq:sec3.2-safe_control_condition_online}, extending the analysis to discrete-time uncertain systems. While Section~3 focuses on safety guarantees, analyzing system behavior under unsafe initial conditions (i.e., $x_0 \notin \cSb{\hat{\para}_0}$) provides another important aspect of the proposed safe control approach. In this regime, the safety filter \eqref{eq:sec3.2-safety_filter} provides a restorative action that steers trajectories toward $\cSr{t}$. Rather than guaranteeing safety, this section establishes asymptotic convergence of $x_t$ to $\cSr{t}$.
\vspace*{-0.2cm}
\begin{definition}
	\label{def:set_sequential_stability}
	Let $\{\cX_t\}^{\infty}_{t=0}$ with $\cX_t \subseteq \cX^\ast$ be a sequentially positively invariant sequence (cf. Definition~\ref{def:invariance_safety_time}) for a general system 
	$x_{t+1} = h_t(x_t) + w_t$, where $h_t: \cX \to \bR^n$ is continuous for all $t \in \obN$, and $w_t \in \cW$. The state $x_t$ is said to converge asymptotically to\footnote{Given a sequence of sets $\{\cX_t\}^{\infty}_{t=0}$, $\limsup_{t\to\infty}\cX_t := \bigcap^{\infty}_{\tau = 0}\left(\bigcup_{t \geq \tau}\cX_t\right)$~\cite{folland1999real}.}$\cX_{\mathrm{sup}} := \limsup_{t\to\infty}\cX_t$ on $\cX^\ast$ if $\lim_{t\to\infty}\dist_{\cX_{\mathrm{sup}}}(x_t) = 0$ for all $x_0 \in \cX^\ast$.
\end{definition}
\vspace*{-0.2cm}
Following \eqref{eq:sec3.2-ra_safe_set}, we define the corresponding \textit{time-varying} continuous energy functions 
on $\cX$ as follows:
\vspace*{-0.2cm}
\begin{equation}
	\label{eq:sec5-candidate_function}
    V_t(x) = 
    \begin{cases}
    0 & \text{if } x \in \cSr{t} \\
    -\Brt{t}(x) & \text{if } x \in \cX \setminus \cSr{t}
    \end{cases}, \quad \forall t \in \obN.
    \vspace*{-0.2cm}
\end{equation}
It is obvious that $\forall t \in \obN, V_t(x) > 0$ for $x \in \cX \setminus \cSr{t}$. Given that $\cSr{t}$ is sequentially positively invariant (see Appendix~\ref{app:proof_1_theorem_2}), $V_t$ can be used to show that $x_t$ converges to the time-varying set $\cSr{t}$ asymptotically if the input $u_t$ satisfies \eqref{eq:sec3.2-CBF_condition_ra}, which is stated in the following theorem.
\vspace*{-0.2cm}
\begin{thm}[Inherent robustness]
\label{thm:robustness}
Given an estimator satisfying Assumption~\ref{ass:bounded_increments_assumption} and a robust adaptive DT-CBF $B: \cX\times \setpara \to \bR$ (cf. Definition~\ref{def:cbf_ra}), then $x_t$ converges asymptotically to $\limsup_{t\to\infty}\cSr{t}$ on $\cX$ (cf. Definition~\ref{def:set_sequential_stability}) for the system \eqref{eq:sec2.2-model_perturbed_true} with $u_t \in \Uasf{t}(x_t;p)$, where $\Uasf{t}(x_t;p)$ is constructed as in \eqref{eq:sec3.2-safe_control_set_online} via~\eqref{eq:sec3.2-safe_control_condition_online}.
\end{thm}
\vspace*{-0.2cm}
The proof of Theorem~\ref{thm:robustness} is provided in Appendix~\ref{app:proof_3_theorem_3}. This result implies that control inputs certified by~\eqref{eq:sec3.2-safe_control_condition_online} guarantee $x_t \in \cSr{t}$ as $t \to \infty$. Although $\cSr{t}$ does not generally converge to $\cS$, asymptotic convergence to the time-varying set $\cSr{t}$ still holds. This property also applies when using the maximized error-bound CBF design \cite{lopez2020robust} discussed in Remark~\ref{rmk:error_choice}. Furthermore, if $\alpha(\cdot)$ is linear or admits a linear lower bound (i.e., $\alpha(r) \ge kr$ for some $k > 0$), exponential convergence is achieved~\cite{bof2018lyapunov}.

%% file: 3-main_content/7-simulation.tex
\vspace*{-0.3cm}
\section{Numerical simulations}
\label{sec7:numerical_example}
\vspace*{-0.3cm}
In this section, two simulation examples are provided. The first one involves vehicle cruise control, a standard benchmark in safe control using CBFs~\cite{taylor2020adaptive}, \cite{xiao2021adaptive}. The second one considers speed control of Permanent-Magnet Synchronous Motors (PMSMs). For both examples, the RLS-based estimator of Section~\ref{sec4:identification} is adopted. The simulations\hspace*{-0.2em}\footnote{The full codebase and specific implementations are available at: \texttt{https://github.com/lcrekko/dt\_ra\_cbf}.}are implemented in Python 3.13.11.
\input{3-main_content/7.1-cruise_control}
\input{3-main_content/7.2-pmsm}

%% file: 3-main_content/7.1-cruise_control.tex
\vspace*{-0.3cm}
\subsection{Vehicle cruise control}
\label{subsec7.1-cruise_control}
\vspace*{-0.3cm}
Applying forward Euler discretization ($\Delta t = 0.1~\si{\second}$) to the continuous-time vehicle model in~\cite{taylor2020adaptive} yields
\vspace*{-0.4cm}
\begin{equation*}
	\label{eq:sec6-acc_model_discrete_time}
	\begin{bmatrix}
		v_{t+1} \\
		d_{t+1}
	\end{bmatrix}
	\seq
	\begin{bmatrix}
		v_t \sminus  \frac{\Delta t}{M}\Fr(v_t)\\
		d_t \splus  \Delta t (\vf \sminus v_t)
	\end{bmatrix}
	\splus
	\begin{bmatrix}
		\frac{\Delta t}{M} \\
		0
	\end{bmatrix}
	u_t
	+
	\begin{bmatrix}
		w_{v,t}\Delta t \\
		w_{d,t}\Delta t
	\end{bmatrix}.
	\vspace*{-0.2cm}
\end{equation*}

\vspace*{-0.6cm}
\noindent where $v$, $M$, and $d$ denote velocity, vehicle mass, and distance to the leading vehicle, respectively. The leading vehicle's nominal velocity $\vf \in [v_{\text{lb}}, v_{\text{ub}}]$ is unknown, and the input $u \in [-u_{\max}, u_{\max}]$ is the ego vehicle's traction force. The resistance force $\Fr(v) = F_{\text{roll}} + \mu_{\text{vis}}v + \mu_{\text{aero}}v^2$ models drag effects for $v > 0$ \cite{taylor2020adaptive}. While $F_{\text{roll}}$ and $\mu_{\text{vis}}$ are determined offline, $\mu_{\text{aero}} \in [\mu^{-}, \mu^{+}]$ is identified online \cite{xiao2021adaptive}. Disturbances $w_{v,t} \in [-\overline{w}_v, \overline{w}_v]$ and $w_{d,t} \in [-\overline{w}_d, \overline{w}_d]$ account for wind effects and variations in $\vf$, respectively. Nominal parameters and known bounds are summarized in Table~1. The model follows the general form \eqref{eq:sec2.2-model} with box-constrained disturbance $w_t = [w_{v,t}\Delta t, w_{d,t}\Delta t]^\top$ and parameter $\theta = [\mu_{\text{aero}}, v_f]^\top$ satisfying \eqref{eq:sec2.2-disturbance_set}--\eqref{eq:sec2.2-parameter_set}. We enforce safety via the time-headway constraint $d \geq 1.8v$ \cite{taylor2020adaptive} using a constructed parameter-free DT-CBF $B(v,d) = d - 1.8v - 0.5$ with $\alpha(x) = (1 - 10^{-4})x$. Numerical verification confirms that \eqref{eq:sec3.2-CBF_condition_ra} holds for $p=2$ and $\Gamma = 10^5\mathbf{I}_2$. For numerical illustration, a myopic (i.e., one-step prediction) adaptive MPC (aMPC) controller serves as the nominal controller, tracking $v_{\text{ref}} = 30$ m/s and utilizing the estimator from Section~\ref{sec4:identification} to obtain estimates $\hat{\mu}_{\text{aero}}$ and $\hat{v}_f$. We compare adaptive MPC with a robust adaptive safety filter (\textbf{aMPC-raCBF}), MPC with a robust safety filter (\textbf{MPC-rCBF}), adaptive MPC (\textbf{aMPC}), and myopic robust adaptive MPC (\textbf{raMPC}) proposed in~\cite{kohler2021robust}. 
\begin{table}[h]
	\label{tab:parameters}
	\centering
	\caption{Parameters and bounds used in cruise control simulation.}
	{\renewcommand{\arraystretch}{0.9}
		\setlength{\tabcolsep}{2pt}
		
		\begin{tabularx}{0.83\columnwidth}{ @{}l c c
				|   
				l c c
				@{}}
			\toprule
			\multicolumn{3}{c|}{\textbf{Nominal parameters}} &
			\multicolumn{3}{c}{\textbf{Bounds}} \\
			\midrule
			Symbol & Value & Unit & 
			Symbol & Value & Unit \\
			\midrule
			$M$ & \num{1650} & \si{\kilogram} &
			$[v_{\text{lb}}, v_{\text{ub}}]$ & [\num{20}, \num{32}] & \si{\metre/\second} \\
			$F_{\text{roll}}$ & \num{125} & \si{\newton} &
			$[\mu^{-}, \mu^{+}]$ & [\num{0.1}, \num{0.6}] & \si{\kilogram/\metre} \\
			$\mu_{\text{vis}}$ & \num{1.2} & \si{\kilogram/\second} &
			$\overline{w}_v$ & \num{2} & \si{\metre/\second\squared} \\
			$\mu^\ast_{\text{aero}}$ & \num{0.55} & \si{\kilogram/\metre} & 
			$\overline{w}_d$ & \num{5} & \si{\metre/\second} \\
			$v^\ast_{\text{f}}$ & \num{22} & \si{\metre/\second} &
			$u_{\max}$ & \num{10000} & \si{\newton} \\
			\bottomrule
		\end{tabularx}
	}
\end{table}
\begin{figure}[h]
	\centering
	\includegraphics[width=0.45\textwidth]{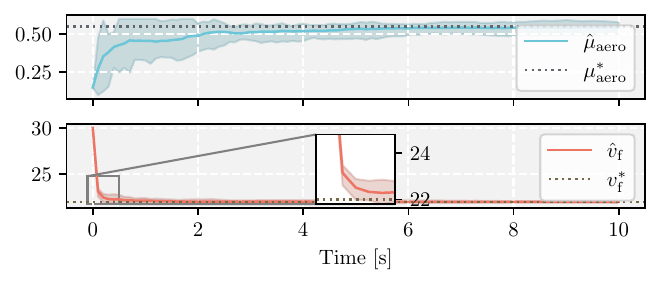}
	\vspace*{-0.3cm}
	\caption{Parameter estimation of vehicle example using unconstrained adaptive MPC with robust adaptive safety filter. The shaded area is the min-max envelope encompassing different disturbance realizations.}
	\label{fig:est_performance}
\end{figure}
\begin{figure}[t]
	\centering
	\includegraphics[width=0.45\textwidth]{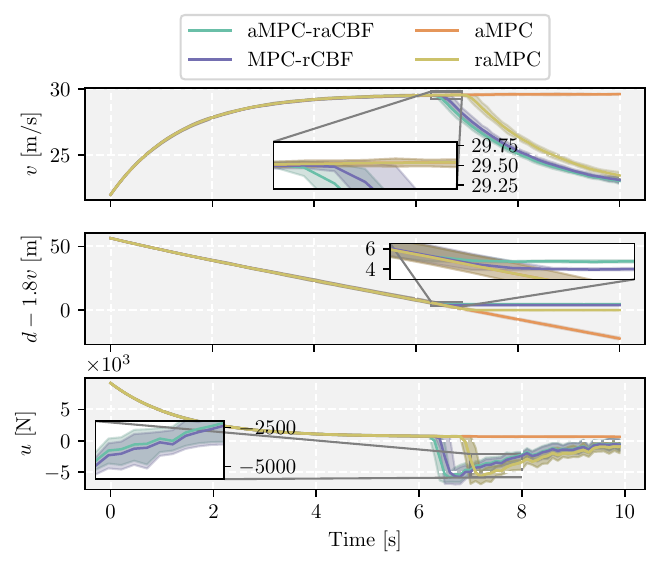}
	\vspace*{-0.3cm}
	\caption{Comparison of adaptive MPC with a robust adaptive safety filter (aMPC-raCBF), MPC with a robust safety filter (MPC-rCBF), adaptive MPC (aMPC), and robust adaptive MPC (raMPC)~\cite{kohler2021robust}. The shaded area denotes the min-max envelope across various disturbance realizations, modeled as uniformly distributed noise.}
	\label{fig:cbf_performance}
\end{figure}
Fig.~\ref{fig:est_performance} shows the performance of the estimator, demonstrating accurate parameter recovery over time. Safety results are summarized in Fig.~\ref{fig:cbf_performance}, where the middle subplot shows that the proposed robust adaptive CBF-based safety filter maintains constraint satisfaction despite disturbances and model mismatch. This is achieved by actively modifying the control inputs as safety limits are approached (Fig.~\ref{fig:cbf_performance}, bottom). Notably, the unconstrained aMPC baseline fails to enforce safety; in this myopic case, it is equivalent to a control Lyapunov function (CLF) approach with an adaptive model. While constrained MPC is a valid alternative for safe control, our purpose is not to outperform constrained MPC but to demonstrate the safety guarantee of CBFs. Indeed, CBFs can be integrated into MPC~\cite{zeng2021enhancing} or used to safeguard controllers that lack inherent guarantees, such as PID control, reinforcement learning~\cite{cohen2023safe}, or approximate MPC~\cite{hertneck2018learning}, \cite{liu2025approximate}. Although robust adaptive MPC with constraint tightening \cite{kohler2021robust} also ensures safety, it reacts more slowly to the time-headway safety requirement in our simulation setting, necessitating more conservative braking beginning at around $t = 7$~s.

%% file: 3-main_content/7.2-pmsm.tex
\vspace*{-0.2cm}
\subsection{Speed control of PMSMs}
\label{subsec6.2:pmsm}
\vspace*{-0.4cm}
Following the surface-mounted PMSM model in~\cite{li2009adaptive} with the direct-axis (d-axis) current $\id$ set to zero, the discrete-time model, under forward Euler discretization with sampling period $\Delta t = 10^{-3}[\si{\second}]$, is given by
\vspace*{-0.3cm}
\begin{equation*}
	\label{eq:sec6-acc_model_discrete_time}
	\hspace*{-0.3ex}
	\begin{bmatrix}
		\omega_{t+1} \\
		\iqt{t+1}
	\end{bmatrix}
	\hspace*{-0.5ex}
	\seq
	\hspace*{-0.5ex}
	\begin{bmatrix}
		\left(1 \sminus \frac{\Bvis\Delta t}{J}\right) & \frac{\np\phif\Delta t}{J} \\
		- \frac{\np\phif\Delta t}{L} & 1\sminus\frac{R\Delta t}{L}
	\end{bmatrix}
	\hspace*{-0.5em}
	\begin{bmatrix}
		\omega_t \\
		\iqt{t}
	\end{bmatrix}
	\splus
	\begin{bmatrix}
		0 \\
		\frac{\Delta t}{L}
	\end{bmatrix}
	\hspace*{-0.3em}
	\uqt{t}
	\splus
	\begin{bmatrix}
		\nu_{\omega,t}\Delta t \\
		\nu_{i,t}\Delta t
	\end{bmatrix}\hspace*{-0.5ex},
\end{equation*}

\vspace*{-0.6cm}
\noindent where $\iq$, $\uq$, and $\omega$ denote the quadrature-axis (q-axis) current, q-axis voltage, and rotor angular velocity, respectively. While the number of pole pairs $\np$, rotor inertia $J$, and inductance $L$ are known, the flux linkage $\phif \in [\phif^-, \phif^+]$, viscous friction coefficient $\Bvis \in [\Bvis^-, \Bvis^+]$, and resistance $R \in [R^-, R^+]$ are \textit{unknown} but bounded. Disturbances $\nu_{\omega,t} \in [-\overline{\nu}_{\omega}, \overline{\nu}_{\omega}]$ and $\nu_{i,t} \in [-\overline{\nu}_i, \overline{\nu}_i]$ account for load variations and perturbations of $\id$, respectively. This discretized PMSM model aligns with \eqref{eq:sec2.2-model} under the box-constrained disturbance $\nu_t = [\nu_{\omega,t}\Delta t, \nu_{i,t}\Delta t]^\top$ and the parameter $\theta = [\phif, \Bvis, R]^\top$. The objective is to track $\omega_{\text{ref}}$ while ensuring the safety constraint $\iq \in [-I_{\max}, I_{\max}]$. To this end, we construct two\footnote{For multiple CBFs, the set in~\eqref{eq:sec3.2-safe_control_set_online} is defined by incorporating the input conditions corresponding to each CBF.}CBFs $B^+(\iq) = I_{\max} - \iq - 0.05$ and $B^-(\iq) = \iq + I_{\max} - 0.05$ with $\alpha(x) = (1 - 10^{-4})x$. Numerical verification confirms that \eqref{eq:sec3.2-CBF_condition_ra} holds for both CBFs with $p=1$ and $\Gamma = 10^7\mathbf{I}_3$. The voltage saturation limit (i.e., bound on $\uq$) is $220\;\si{\volt}$. Nominal parameters and bounds are listed in Table~2. We use an adaptive PID (aPID) controller as the nominal controller and compare aPID control with a robust adaptive safety filter (\textbf{aPID-raCBF}), aPID with a robust safety filter (\textbf{aPID-rCBF}), PID control with a robust safety filter (\textbf{PID-rCBF}), and input-saturated \textbf{aPID} control.
\begin{figure}[h]
	\centering
	\includegraphics[width=0.45\textwidth]{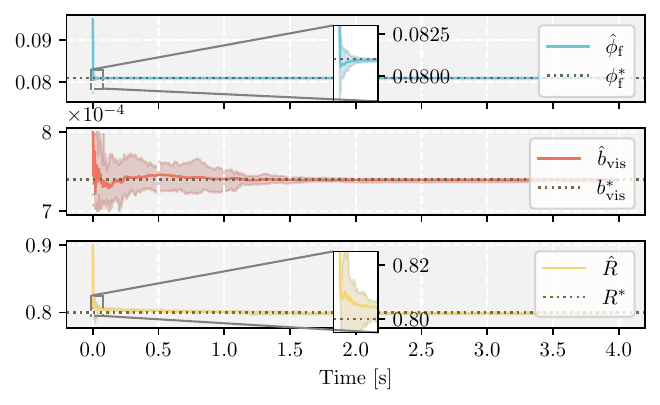}
	\vspace*{-0.3cm}
	\caption{Parameter estimation of PMSM example using adaptive PID control with robust adaptive safety filter. The shaded area is the min-max envelope across various disturbance realizations, modeled as uniformly distributed noise.}
	\label{fig:est_performance_pmsm}
\end{figure}
\begin{figure}[t]
	\centering
	\includegraphics[width=0.45\textwidth]{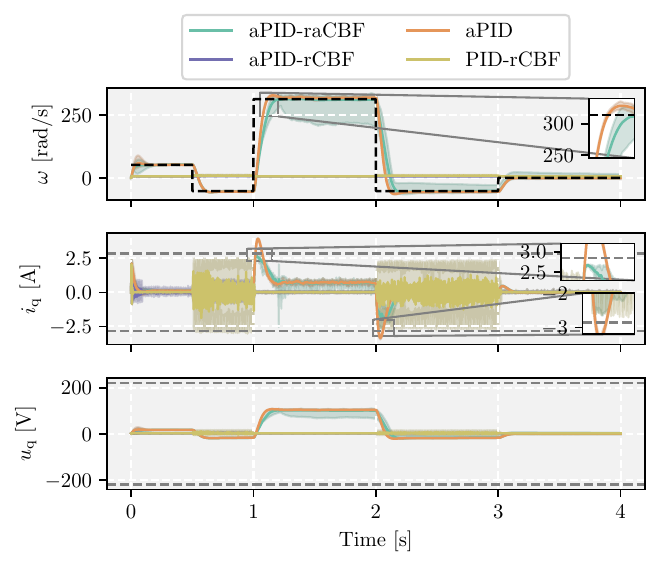}
	\vspace*{-0.3cm}
	\caption{Comparison of adaptive PID (aPID) control with a robust adaptive safety filter (aPID-raCBF), aPID with a robust safety filter (aPID-rCBF), PID control with a robust safety filter (PID-rCBF), and input-saturated aPID control. The shaded area denotes the min-max envelope across various disturbance realizations, modeled as uniformly distributed noise.}
	\label{fig:cbf_performance_pmsm}
\end{figure}
\begin{table}[h]
	\label{tab:parameters_pmsm}
	\centering
	\caption{Parameters and bounds used in motor control simulation.}
	{\renewcommand{\arraystretch}{0.9}
		\setlength{\tabcolsep}{2pt}
		
		\begin{tabularx}{0.98\columnwidth}{ @{}l c c
				|   
				l c c
				@{}}
			\toprule
			\multicolumn{3}{c|}{\textbf{Nominal parameters}} &
			\multicolumn{3}{c}{\textbf{Bounds}} \\
			\midrule
			Symbol & Value & Unit & 
			Symbol & Value & Unit \\
			\midrule
			$\np$ & \num{4} & \si{[-]} &
			$[\phif^{-}, \phif^{+}]$ & $[0.065, 0.095]$ & \si{\weber} \\
			$J$ & \num{2.35e-4} & \si{\kilogram\metre\squared} &
			$[\Bvis^{-}, \Bvis^{+}]$ & $[7, \;8]\cdot10^{-4}$ & \si{\newton\metre\second}\\
			$L$ & \num{2.9e-3} & \si{\henry} &
			$[R^{-}, R^{+}]$ & $[0.6,\; 1]$ & \si{\ohm} \\
			$\phif^\ast$ & \num{0.081} & \si{\weber} & 
			$\overline{v}_{\omega}$ & $100$ & \si{\per\second\squared} \\
			$\Bvis^\ast$ & \num{7.42e-4} & \si{\newton\metre\second}&
			$\overline{v}_i$ & \num{60} & \si{\ampere/\second} \\
			$R^\ast$ & \num{0.8} & \si{\ohm} &
			$I_{\max}$ & \num{2.8} & \si{\ampere} \\
			\bottomrule
		\end{tabularx}
	}
\end{table}

\vspace*{-0.5cm}
Fig.~\ref{fig:est_performance_pmsm} illustrates the high estimation accuracy achieved for all three unknown parameters. The comparative efficacy of the proposed framework is illustrated in Fig.~\ref{fig:cbf_performance_pmsm}. While the baseline \textbf{aPID} controller achieves precise tracking, it fails to maintain safety as the q-axis current violates the prescribed bounds. Conversely, the \textbf{aPID-rCBF} and \textbf{PID-rCBF} variants ensure safety but exhibit extreme conservatism; because the parameter error $\bet{t}(p)$ is not reduced via estimation, the robustification term $\|\bm{\phi}(x)\|\bet{t}(p)$ severely restricts control effort, resulting in a near-total loss of tracking. In contrast, the proposed robust adaptive safety filter overcomes this issue. By dynamically updating uncertainty sets, it maintains rigorous safety certificates with only marginal tracking degradation, effectively achieving satisfactory tracking performance while maintaining robust safety.

%% file: 3-main_content/8-conclusion.tex
\vspace*{-0.2cm}
\section{Conclusions and future work}
\label{sec8:conclusion}
\vspace*{-0.3cm}
In this paper, we have proposed a robust adaptive safe control method for discrete-time systems under parametric uncertainty and disturbances. Unlike traditional continuous-time designs that couple adaptation and control, our framework introduces a modular architecture where the parameter estimator operates independently of the safety filter. Future work will focus on extending this framework to output-feedback systems with imperfect measurements and/or nonlinearly parameterized uncertainty. Additionally, we intend to investigate active exploration for enhanced estimation, predictive safety filter formulations, the formal synthesis of control barrier functions (CBFs) for uncertain systems, and safe control with finite-time robustness guarantees.

%% file: 6-appendices/0-app_main.tex
\appendix
\input{6-appendices/1-thm2_proof}
\input{6-appendices/2-subtlety}

\input{6-appendices/3-thm3_robustness}

%% file: 6-appendices/1-thm2_proof.tex
\vspace*{-0.4cm}
\section{Proof of Theorem \ref{thm:cbf_ra}}    
\vspace*{-0.7cm}
\label{app:proof_1_theorem_2}
\begin{pf}
	First, due to the definition of $\Delta B^\star(x,u)$ and $B^{\star}(x)$, the increment bound in Assumption~\ref{ass:bounded_increments_assumption}, and $\bar{\varepsilon}_{\para}(p) \geq \varepsilon_{\para,t}(p)$, it can be obtained that \eqref{eq:sec3.2-CBF_condition_ra} is a sufficient condition for \eqref{eq:sec3.2-safe_control_condition_online}. Therefore, if $B$ is a robust adaptive DT-CBF that satisfies \eqref{eq:sec3.2-CBF_condition_ra}, then $\Uasf{t}(x;p)$ is non-empty. Next, it suffices to prove robust safety through positive invariance. Considering $\Brt{t+1}(x_{t+1}) - \Brt{t}(x_t)$, which characterizes the increment of the barrier function, it holds that
	\vspace*{-0.3cm}
	\begin{align*}
		  & \Brt{t+1}(x_{t+1}) - \Brt{t}(x_t) \notag \\
		& \hspace*{-3ex} \overset{\eqref{eq:sec3.1-estimate_increment}}{=} B(x_{t+1},\hpa{t+1}) - B(x_t,\hpa{t}) - \frac{1}{2}\ipa{t}^\top \Gamma^{-1}(\ipa{t} + 2\epat{t}) \notag \\
		\geq & B(x_{t+1},\hpa{t+1}) - B(x_t,\hpa{t}) - \frac{\|\ipa{t}\|^2 + 2\|\ipa{t}\|\bet{t}(p)}{2\mG},
		\vspace*{-0.5cm}
	\end{align*}
	where $B(x_{t+1},\hpa{t+1}) - B(x_t,\hpa{t})$ can be further relaxed using the Lipschitz perturbation bound as
	\vspace*{-0.3cm}
	\begin{multline*}
		  B(x_{t+1},\hpa{t+1}) - B(x_t,\hpa{t})
		\geq B(f(x_t,\pi_t(x_t);\hpa{t}),\hpa{t}) \\ - B(x_t,\hpa{t}) -L_{B,x}(\|\bm{\phi}(x_t)\|\bet{t}(p) +  \overline{w}) - L_{B,\para}\|\ipa{t}\|.
	\end{multline*}
	Combining the above two inequalities and using the fact that $\pi_t(x_t)$ satisfies \eqref{eq:sec3.2-safe_control_condition_online}, we have
	\vspace*{-0.25cm}
	\begin{multline}
		\label{eq:app1-final_relation}
		\hspace*{-2ex}\Brt{t+1}(x_{t+1}) - \Brt{t}(x_t) \\ \hspace*{-4ex}\geq -\alpha\left(B(x_t,\hpa{t}) \sminus \frac{(\bet{t}(p))^2}{2\mG}\right) \geq -\alpha(\Brt{t}(x_t)).
	\end{multline}
	Based on \eqref{eq:app1-final_relation}, if $x_t \in \cSr{t}$, i.e., $\Brt{t}(x_t) \geq 0$, then $\Brt{t+1}(x_{t+1}) \geq \Brt{t}(x_t) \sminus \alpha(\Brt{t}(x_t)) \geq 0$ since $\alpha \in \csl$, which implies $x_{t+1} \in \cSr{t+1}$. Thus, $\{\cSr{t}\}^{T}_{t=0}$ is positively invariant for the closed-loop system \eqref{eq:sec3.2-closed_theorem_example} given $\pi_t(x) \in \Uasf{t}(x;p)$. Moreover, noting that $\Brt{t}(x) \geq 0 \Longrightarrow B(x,\hpa{t}) \geq 0$, then $\{\cSb{\hpa{t}}\}^{T}_{t=0}$ is positively invariant (i.e., robustly safe) for \eqref{eq:sec3.2-closed_theorem_example}. \hspace*{19ex}\qed
\end{pf}

%% file: 6-appendices/2-subtlety.tex
\vspace*{-0.6cm}
\section{Robustified CBFs with error bounds}    
\label{app:subtlety}
\vspace*{-0.3cm}
The true error $\epat{t}$ is used to construct $\Brt{t}$ in \eqref{eq:sec3.2-ra_barrier_function} to achieve adaptive safety, which is similar to \cite{taylor2020adaptive}. Although safety can be established, $\{\cSr{t}\}^{T}_{t=0}$ cannot be characterized since $\epat{t}$ is unknown. Considering that the ultimate goal is to achieve safety w.r.t.~$\cSb{\hpa{t}}$, which encodes safety \cite{lopez2020robust}, \cite{taylor2020adaptive}, the proposed method using $\Brt{t}$ defined in \eqref{eq:sec3.2-ra_barrier_function} is still valid since $\cSr{t} \subseteq \cSb{\hpa{t}}$. 

\vspace*{-0.4cm}
An alternative is to use the \textit{set-induced} error bound $\eet{t}$ satisfying $|\epat{t}| \leq \eet{t}$ to construct robustified CBFs \cite{lopez2020robust}. Specifically, for all $i \in \bI_{[1:q]}$, $\eet{t}[i]$ can be computed as
\vspace*{-0.3cm}
\begin{equation}
	\label{eq:app2-error_bound_set_entry}
	\eet{t}[i] = \max_{\theta \in \setpara_t}\theta[i] - \min_{\theta \in \setpara_t}\theta[i].
	\vspace*{-0.2cm}
\end{equation}
The bound in \eqref{eq:app2-error_bound_set_entry} is conservative as it depends solely on $\setpara_t$. Furthermore, the non-smoothness of the $\max$ and $\min$ operations precludes an analytical time derivative, forcing continuous-time invariance proofs to rely on the true error rather than the bound used to construct the CBF \cite{lopez2020robust}. Consequently, establishing the invariance of safe sets defined by error-bound robustified CBFs is generally infeasible in continuous time\footnote{In the proof of \cite[Theorem 2]{lopez2020robust}, the conditions $h \geq h_r - \frac{1}{2}\tilde{\vartheta}^\top\Gamma^{-1}\tilde{\vartheta}$ and $h \geq 0$ do not imply $h_r - \frac{1}{2}\tilde{\vartheta}^\top\Gamma^{-1}\tilde{\vartheta} \geq 0$.}\hspace*{-0.2em}. In contrast, the discrete-time setting permits the establishment of such invariance properties as no differentiation is required. The error-bound robustified CBF is defined as:
\vspace*{-0.3cm}
\begin{equation}
	\label{eq:app2-barrier_function}
	\BBrt{t}(x) := B(x,\hpa{t}) - \frac{1}{2}\eet{t}^\top\Gamma^{-1}\eet{t},
	\vspace*{-0.2cm}
\end{equation}
where $\Gamma$ is a symmetric positive definite matrix, and the safe set corresponding to $\BBrt{t}$ is given by
\vspace*{-0.3cm}
\begin{equation}
	\label{eq:app2-ra_safe_set}
	\bcSr{t} = \{x \in \cX \mid \BBrt{t}(x) \geq 0\}.
	\vspace*{-0.2cm}
\end{equation}
Following Definition \ref{def:cbf_ra}, a modified robust adaptive DT-CBF based on \eqref{eq:app2-barrier_function} and \eqref{eq:app2-ra_safe_set} can be defined. Moreover, a safety guarantee similar to Theorem~\ref{thm:cbf_ra} follows.
\vspace*{-0.2cm}
\begin{definition}
	\label{def:app2-cbf_ra}
	Given a set-based estimator $\est$, a continuous function $B: \cX \times \setpara \to \bR$ is an error-bound robust adaptive DT-CBF if it is Lipschitz continuous and there exists $\alpha \in \ke \cap \mnt{\csi}$ such that, for all $x \in \cX$, there exists $u \in \cU$ satisfying
	\vspace*{-0.4cm}
	\begin{multline}
		\label{eq:app2-CBF_condition_ra}
		\Delta B^\star(x,u) - L_{B,x}\overline{w} - E^{\star}_{\eta}(x) \\ \geq -\alpha\left(B^{\star}(x) - \frac{\|\eta_{\para,0}\|^2}{2\mG} \right),
	\end{multline}
	
	\vspace*{-0.6cm}
	\noindent where $\Gamma$ is given in~\eqref{eq:app2-barrier_function}, $\Delta B^\star(x,u)$ and $B^{\star}(x)$ are defined in Definition~\ref{def:cbf_ra}, and $E^{\star}_{\eta}(x) \seq \left(L_{B,x}\|\bm{\phi}(x)\| + L_{B,\para}\right)$ $\|\eet{0}\|$ with $L_{B,x}$ and $L_{B,\para}$ given in Definition \ref{ass:lipschitz_base_cbf}. The condition \eqref{eq:app2-CBF_condition_ra} is called the error-bound robust adaptive CBC against disturbances and parametric model mismatch.
\end{definition}
\vspace*{-0.2cm}
\begin{thm}
	\label{thm:app2-cbf_ra}
	Given $x_0 \in \cSr{0}$, a horizon $T \in \bN$, a set-based estimator $\est$, and an error-bound robust adaptive DT-CBF with its associated matrix $\Gamma$ and function $\alpha \in \ke \cap \mnt{\csi}$, for the error-bound robust adaptive online input condition
		\vspace*{-0.4cm}
		\begin{multline}
			\label{eq:app-safe_control_condition_online}
			B\big(f(x,u;\hpa{t}),\hpa{t}\big) - B(x,\hpa{t}) - L_{B,x}\overline{w} \\ -E_{\eta,t}(x)  \geq -\alpha\left(B(x,\hpara_t) - \frac{\|\eta_{\para,t}\|^2}{2\mG}\right),
			\vspace*{-0.2cm}
		\end{multline}
		
		\vspace*{-0.7cm}
		\noindent the admissible input set $\BUasf{t}(x)$ defined as
		\vspace*{-0.3cm}
		\begin{equation}
			\label{eq:app-safe_control_set_online}
			\BUasf{t}(x) := \left\{u \in \cU \mid \eqref{eq:app-safe_control_condition_online} \text{ holds}\right\}
			\vspace*{-0.2cm}
		\end{equation}
		is non-empty, $\forall x \in \cX$ and $t \in \bI_{[0,T-1]}$. The term $E_{\eta,t}(x)$ in \eqref{eq:app-safe_control_condition_online} equals $L_{B,x}\|\bm{\phi}(x)\|\|\eta_{\para,t}\| +L_{B,\theta}\|\delta_{\eta,t}\|$ with $\delta_{\eta,t} = \eet{t+1} - \eet{t}$. In addition, if a time-varying policy $\overline{\pi}_t: \cX \to \bR^m$ satisfy $\overline{\pi}_t(x) \in \BUasf{t}(x)$ for all $x \in \cX$ and $t \in \bI_{[0,T-1]}$, then the closed-loop system
	\vspace*{-0.3cm}
	\begin{equation}
		\label{eq:app2-closed_theorem_example}
		x_{t+1} = f(x_t,\overline{\pi}_t(x_t);\tpara) + w_t
		\vspace*{-0.2cm}
	\end{equation}
	is robustly safe w.r.t.~$\{\bcSr{t}\}^{T}_{t=0}$ and thus robustly safe w.r.t.~$\{\cSb{\hpara_t}\}^{T}_{t=0}$ (cf. Definition \ref{def:invariance_safety_time}) against both parametric model mismatch and disturbances.
\end{thm}

The proof of Theorem~\ref{thm:app2-cbf_ra} is similar to that of Theorem~\ref{thm:cbf_ra} in Appendix~\ref{app:proof_1_theorem_2} and is therefore omitted for brevity.
\vspace*{-0.2cm}
\begin{remark}
	\label{rmk:monotonicity}
	The advantage of using $\eet{t}$ as in \eqref{eq:app2-barrier_function} is that $\BBrt{t}$ can be explicitly determined. Moreover, using set-membership identification, it holds that $\eet{t+1} \leq \eet{t}$ (i.e., the error does not increase) since $\setpara_{t+1} \subseteq \setpara_t$, leading to $\bcSr{t} \subseteq \bcSr{t+1}$, i.e., the robust adaptive safe set defined using the error bound is non-shrinking.
\end{remark}

%% file: 6-appendices/3-thm3_robustness.tex
\vspace*{-0.3cm}
\section{More on inherent robustness}
\vspace*{-0.3cm}
\label{app:proof_3_theorem_3}
We first prove Theorem~\ref{thm:robustness} using $V_t$ defined in \eqref{eq:sec5-candidate_function}. The proof is inspired by the proof of \cite[Theorem~B.13]{rawlings2017model}.
\vspace*{-0.7cm}
\begin{pf}
	If there exists a $t_0 \in \bN$ such that $x_{t_0} \in \cSr{t_0}$, then $x_t \in \cSr{t}$ holds for all $t \in \bI_{[t_0:\infty]}$ by Theorem~\ref{thm:cbf_ra} since $\{\Brt{t}\}^{\infty}_{t=0}$ is sequentially positively invariant for the system \eqref{eq:sec2.2-model_perturbed_true} with $u_t \in \Uasf{t}(x_t;p)$. Thus, in this case, $\lim_{t\to\infty}\dist_{\cSr{t}}(x_t) = 0$. On the other hand, assume $x_t \in \cX \setminus \cSr{t}$; then we have
	\vspace*{-0.35cm}
	\begin{multline}
		\label{eq:app3-robustness_core}
		V_{t+1}(x_{t+1}) - V_t(x_t) \overset{\eqref{eq:sec5-candidate_function}}{=} \Brt{t}(x_t) - \Brt{t+1}(x_{t+1}) \\ \overset{\eqref{eq:app1-final_relation}}{\leq} \alpha(\Brt{t}(x_t)) < 0,
		\vspace*{-0.2cm}
	\end{multline}
	where the last inequality is due to $\Brt{t}(x_t) < 0$ for $x_t \in \cX \setminus \cSr{t}$ and $\alpha \in \kei$. Consider the sequence $\{V_t(x_t)\}^{\infty}_{t=0}$; given that $V_t(x_t) \geq 0$ and \eqref{eq:app3-robustness_core}, by the monotone convergence theorem, there exists a $V^\ast \geq 0$ such that $\lim_{t\to\infty}V_t(x_t) = V^\ast$. Equivalently, $\forall \epsilon > 0$, $\exists T^\ast(\epsilon)$ such that $\forall t \geq T^\ast(\epsilon)$, $V_t(x_t) - V^\ast < \epsilon$. Next, we prove that $V^\ast = 0$ by contradiction. Assume $V^\ast > 0$, and let $\epsilon = -\alpha(-V^\ast)$. Then, for any $t \geq T^\ast(-\alpha(-V^\ast))$, it holds that (i) $V^\ast \leq V_{t}(x_{t}) < V^\ast - \alpha(-V^\ast)$ and (ii) $V^\ast \leq V_{t+1}(x_{t+1}) < V^\ast - \alpha(-V^\ast)$, implying $V_{t+1}(x_{t+1}) - V_{t}(x_{t}) > \alpha(-V^\ast)$. On the other hand, by \eqref{eq:app3-robustness_core}, we have $V_{t+1}(x_{t+1}) - V_{t}(x_{t}) \leq \alpha(-V^\ast)$, which leads to a contradiction. Therefore, $\lim_{t\to\infty}V_t(x_t) = V^\ast = 0$, implying $\lim_{t\to\infty}\dist_{\cSr{t}}(x_t) = 0$ by the continuity of $V_t$. \hspace*{3ex}\qed
\end{pf}
\vspace*{-0.5cm}
In fact, following the discussion in Appendix~\ref{app:subtlety}, by constructing a sequence of energy functions using $\bcSr{t}$, it can be shown that the safe control input characterized by the certificate \eqref{eq:app2-CBF_condition_ra} also has the inherent robustness property, which is summarized in the following theorem:
\vspace*{-0.2cm}
\begin{thm}
	\label{thm:app3-robustness}
	Given a set-based estimator and an error-bound robust adaptive DT-CBF $B: \cX\times \setpara \to \bR$ (cf. Definition~\ref{def:app2-cbf_ra}), $x_t$ converges to $\limsup_{t\to\infty}\bcSr{t}$ 
	asymptotically (cf. Definition~\ref{def:set_sequential_stability}) on $\cX$ for the system \eqref{eq:sec2.2-model_perturbed_true} with $u_t \in \BUasf{t}(x_t)$, where $\BUasf{t}(x_t)$ is constructed as in \eqref{eq:app-safe_control_set_online} via~\eqref{eq:app-safe_control_condition_online}.
\end{thm}
\vspace*{-0.2cm}
Theorem~\ref{thm:app3-robustness} implies that $x_t$ will be regulated towards the set $\limsup_{t\to\infty}\bcSr{t}$ asymptotically on $\cX$ for the closed-loop system \eqref{eq:app2-closed_theorem_example}. The proof of Theorem~\ref{thm:app3-robustness} closely follows that of Theorem~\ref{thm:robustness} and is omitted for brevity.

%% file: 3-main_content/9-biography.tex
\vspace*{-0.3cm}
\section*{Biographies}

\begin{figure}[!h]
	\centering
	\begin{minipage}{0.22\linewidth}
		\centering
		\includegraphics[width=\linewidth]{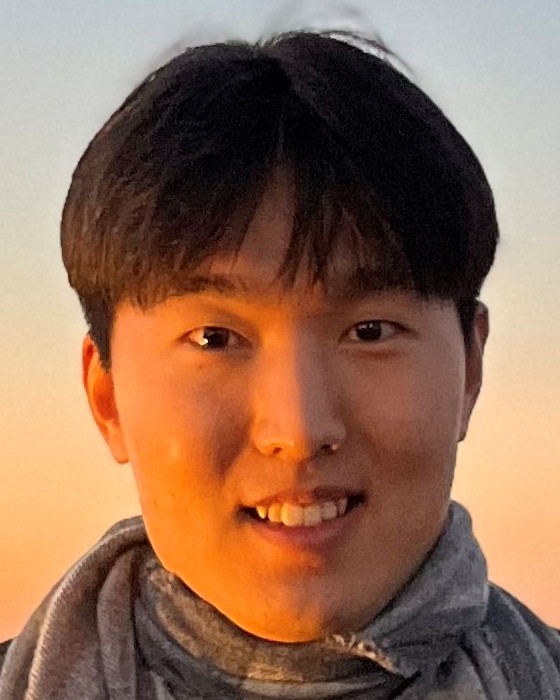}
	\end{minipage}\hfill
	\begin{minipage}{0.75\linewidth}
		{\scriptsize \textbf{Changrui Liu} received the B.Sc. degree (cum laude \& Best 10 Graduates) from the Honors School of Harbin Institute of Technology (HIT), Harbin, China, in 2019, and the M.Sc. degree from TU Delft, Delft, The Netherlands, in 2022. He is currently pursuing the Ph.D. degree at the Delft Center for Systems and Control, TU Delft. His research interests include performance guarantees in machine learning and optimization.}
	\end{minipage}
\end{figure}
\vspace*{0.5em}
\begin{figure}[!h]
	\centering
	\begin{minipage}{0.22\linewidth}
		\centering
		\includegraphics[width=\linewidth]{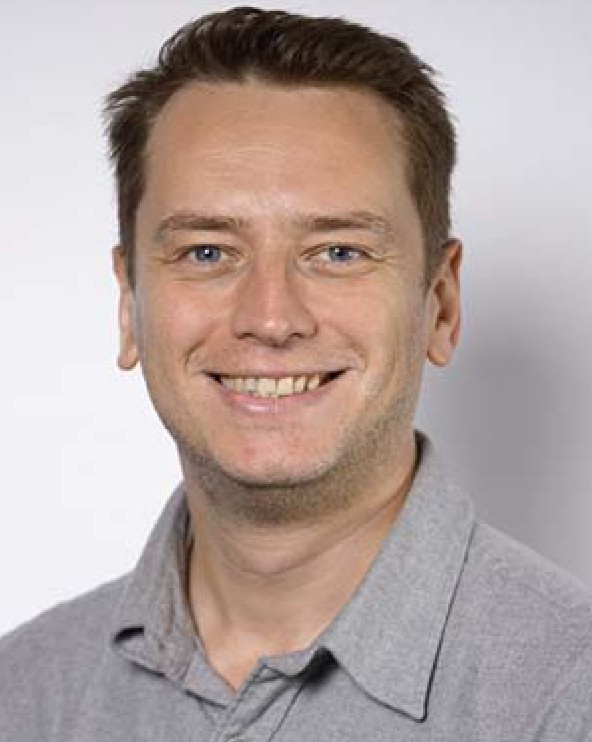}
	\end{minipage}\hfill
	\begin{minipage}{0.75\linewidth}
		{\scriptsize \textbf{Anil Alan} received the
			B.Sc. degree of Mechanical Engineering at Middle East Technical University, Ankara, Türkiye, in 2012, the M.Sc. degree from Bilkent University, Ankara, in 2017, and the Ph.D. degree from the University of Michigan, Ann Arbor, MI, USA, in 2024. He is currently a Postdoc with the Delft University of Technology, Delft, The Netherlands. His research focuses on safety-critical control.}
	\end{minipage}
\end{figure}
\vspace*{0.5em}
\begin{figure}[!h]
	\centering
	\begin{minipage}{0.22\linewidth}
		\centering
		\includegraphics[width=\linewidth]{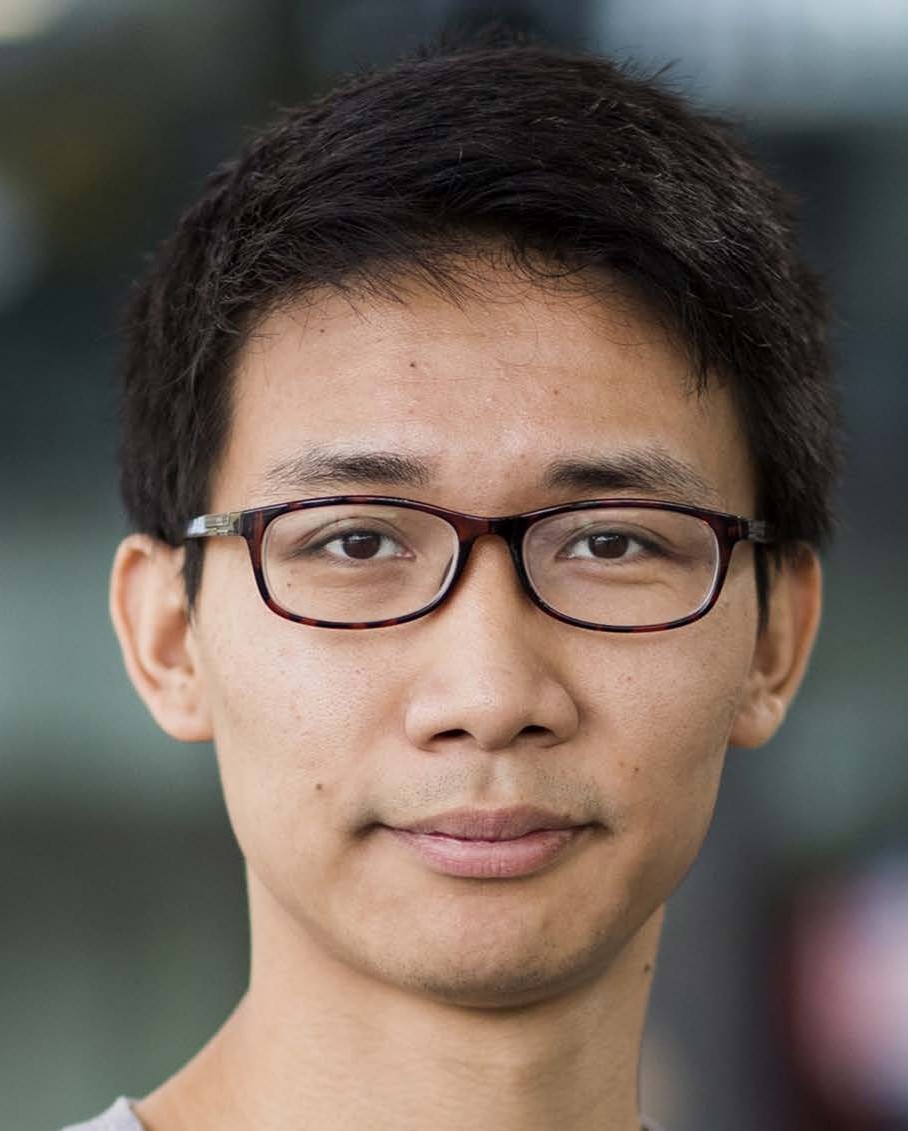}
	\end{minipage}\hfill
	\begin{minipage}{0.75\linewidth}
		{\scriptsize \textbf{Shengling Shi} is an assistant professor at the Delft Center for Systems and Control (DCSC), TU Delft, and he was a Postdoc at the Massachusetts Institute of Technology. He was also a postdoc at the DCSC, TU Delft. He received the Ph.D. degree from the Eindhoven University of Technology in 2021. His research interests include system identification, model predictive control, and their applications.}
	\end{minipage}
\end{figure}
\vspace*{0.5em}
\begin{figure}[!h]
	\centering
	\begin{minipage}{0.22\linewidth}
		\centering
		\includegraphics[width=\linewidth]{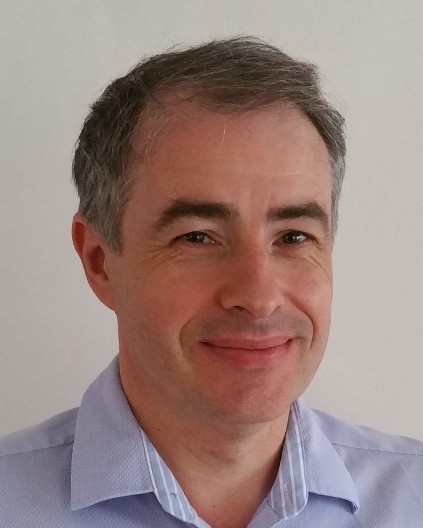}
	\end{minipage}\hfill
	\begin{minipage}{0.75\linewidth}
		{\scriptsize \textbf{Bart De Schutter} is a full professor at the Delft Center for Systems and Control of TU Delft. His research interests include integrated learning- and optimization-based control, machine learning for control and decision making, multi-agent control, and control of hybrid systems with applications in intelligent transportation, smart energy, and underwater robotics.}
	\end{minipage}
\end{figure}